\documentclass[journal]{IEEEtran}

\usepackage{float} \restylefloat{figure} 
\usepackage{color} 

\usepackage{graphicx}
\graphicspath{{eps/}}

\usepackage{fixltx2e}
\usepackage{epstopdf}
\usepackage{rotating}
\usepackage{lscape}
\usepackage{bm} 
\usepackage{cite}
\usepackage{url}
\usepackage[small]{caption}
\usepackage{subfig} 
\usepackage{array} 
\usepackage{mathtools}

\usepackage{amsmath}   
\usepackage{amsfonts}
\usepackage{subfloat}
\usepackage{afterpage}
\long\def\symbolfootnote[#1]#2{\begingroup\def\thefootnote{\fnsymbol{footnote}}
\footnote[#1]{#2}\endgroup}

\begin{document}

\title{{\Large \textbf{Liquid State Machine with Dendritically Enhanced Readout for Low-power, Neuromorphic VLSI Implementations}}}

\author{\authorblockN{Subhrajit Roy,~\IEEEmembership{Student Member,~IEEE}, Amitava Banerjee and Arindam Basu,~\IEEEmembership{Member,~IEEE}}\\
\thanks{The authors are with the School of Electrical and Electronic Engineering,
Nanyang Technological University, Singapore 639798.(e-mail:arindam.basu@ntu.edu.sg). This work was supported by MOE through grants RG 21/10 and ARC 8/13.}
 }
\maketitle
\begin{abstract}
In this paper, we describe a new neuro-inspired, hardware-friendly readout stage for the liquid state machine (LSM), a popular model for reservoir computing. Compared to the parallel perceptron architecture trained by the \emph{p}-delta algorithm, which is the state of the art in terms of performance of readout stages, our readout architecture and learning algorithm can attain better performance with significantly less synaptic resources making it attractive for VLSI implementation. Inspired by the nonlinear properties of dendrites in biological neurons, our readout stage incorporates neurons having multiple dendrites with a lumped nonlinearity (two compartment model). The number of synaptic connections on each branch is significantly lower than the total number of connections from the liquid neurons and the learning algorithm tries to find the best `\emph{combination}' of input connections on each branch to reduce the error. Hence, the learning involves network rewiring (NRW) of the readout network similar to structural plasticity observed in its biological counterparts. We show that compared to a single perceptron using analog weights, this architecture for the readout can attain, even by using the same number of binary valued synapses, up to $3.3$ times less error for a two-class spike train classification problem and $2.4$ times less error for an input rate approximation task. Even with $60$ times larger synapses, a group of $60$ parallel perceptrons cannot attain the performance of the proposed dendritically enhanced readout. An additional advantage of this method for hardware implementations is that the `choice' of connectivity can be easily implemented exploiting address event representation (AER) protocols commonly used in current neuromorphic systems where the connection matrix is stored in memory. Also, due to the use of binary synapses, our proposed method is more robust against statistical variations.
\end{abstract}


\IEEEpeerreviewmaketitle
\section{Introduction and Motivation}
Spiking neural networks, often referred to as the third generation of neural networks, are known to be more bio-realistic and computationally powerful than their predecessors. Since the neuronal communication is in the form of noise-robust, digital pulses or `spikes', these networks are also amenable for low-power, low-voltage very large scale integrated circuit (VLSI) implementations. Hence, in parallel to the progress in theoretical studies of spiking neurons, neuromorphic engineers have been developing low-power VLSI circuits that emulate sensory systems \cite{AEREAR,RALPH,IMAGER} and higher cognitive functions like learning and memory. With the advent of brain-machine interfaces, there is also the need for ultra-low power spike train classifiers that can be used to decode, for example, motor intentions \cite{NITISH1,NITISH2}. One of the major problems in using current neuromorphic designs for practical machine learning problems is the requirement for high-resolution, non-volatile, tunable synaptic weights \cite{Hungyi2012,mill2011,Koickal2007}. An architectural solution is partly provided by the Liquid State Machine (LSM) \cite{Maass2002}, which requires training of very few weights while the others can be random. 
\begin{figure}[t]
\begin{center}
\includegraphics[width=0.4\textwidth]{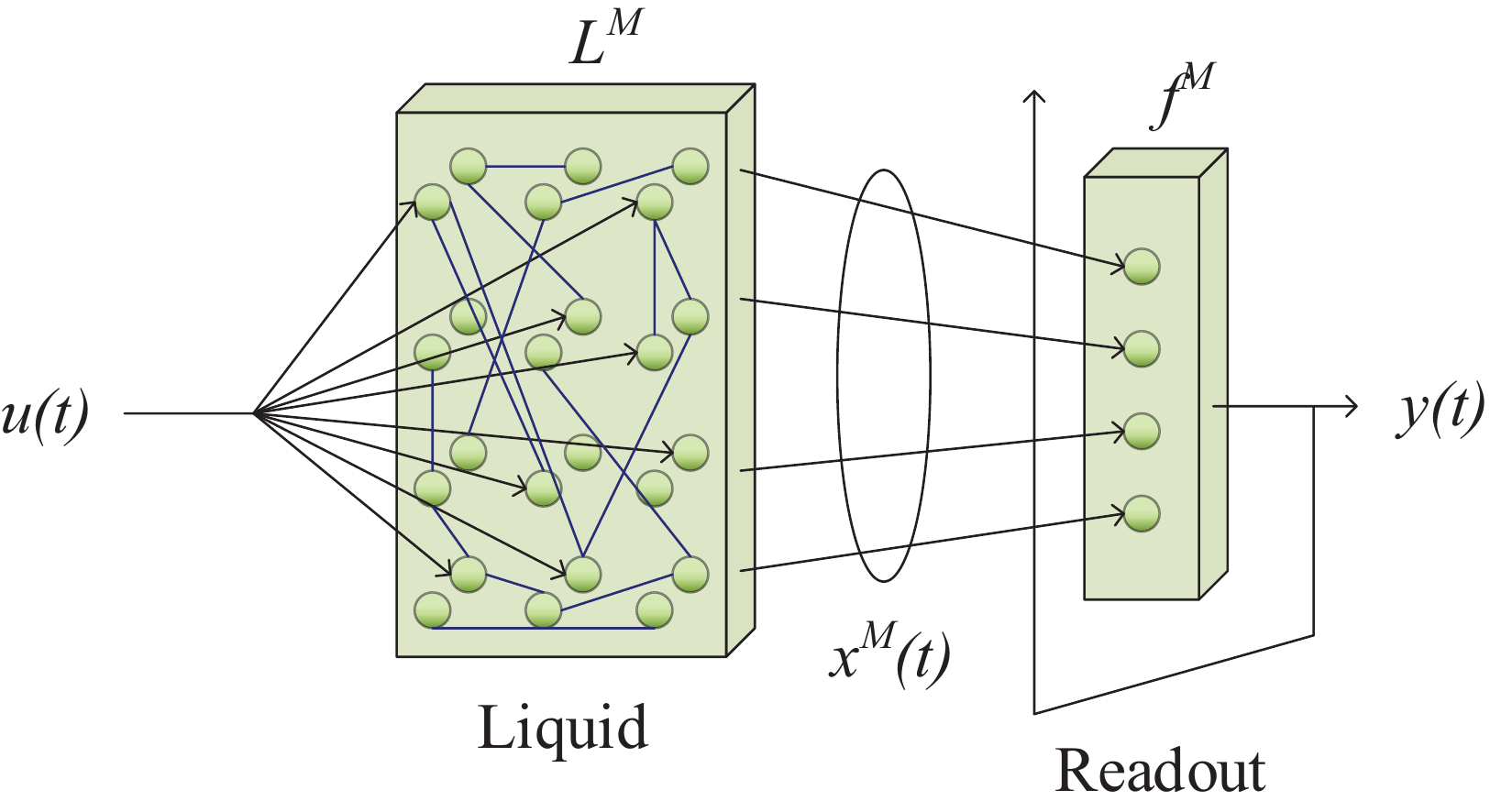}
\caption{The first stage of LSM is the input layer. The input stage is followed by a pool of recurrent LIF neurons whose synaptic connections are not trained. The next stage is a simple linear classifier that is selected and trained in a task-specific manner.}
 \label{fig:LSM}
\end{center}
\end{figure}

The LSM network, depicted in Fig.\ref{fig:LSM}, consists of three stages: an input layer which projects the input pattern into the second stage or the liquid, which is a recurrent neural network (RNN) with randomly weighted, mostly local interconnections performing the task of mapping the input to internal states. These states are then used by the third stage or the readout circuit to provide the overall system output. The readout is trained in a task-specific way which requires updating the weights of the synapses connecting the liquid and the readout. Hence, this does not entirely eliminate the need for high-resolution, non-volatile, tunable weights.

RNNs in general, though computationally very powerful, need updating of the weights of the synapses forming the network. Though several learning techniques for modifying the synaptic weights have been proposed \cite{Werbos,Atiya00newresults,Puskorius94,MAJI,Hochreiter:1997}, an optimal solution is yet to be found. In this context, LSM provides a big advantage since only the weights of the synapses connecting the RNN to the readout need to be modified in an LSM, while the interconnections within the RNN pool are fixed. This is clearly useful for VLSI implementations since tunability or high resolution are not required for most weights. However, the number of tunable weights needed in the readout stage for good performance may become a bottleneck. The state of the art readout stage of LSM using a single layer network as readout, is usually composed of a single layer of parallel perceptrons (we denote LSM with a parallel perceptron readout as LSM-PPR) that are trained by the \emph{p}-delta algorithm \cite{pdelta,Maass2002,zhang2009,raey2011,lengenstein2003}. For this readout stage of LSM-PPR, the number of tunable weights will be very high and equal to ( $L\times n$ ), where $L$ and $n$ are the number of liquid and readout neurons respectively, thereby making it infeasible for low-power smart sensors. To decrease the number of tunable synapses, neuromorphic systems often use an asynchronous multiplexing technique called address event representation (AER) where the connection matrix is stored in a configurable digital memory. Using AER, it is possible to have a large number of synapses for readout; however, the huge power dissipated in accessing memory for every spike makes this solution infeasible for low-power applications.

In this article, we propose a novel architecture and training procedure for the readout stage of LSM that is inspired by the nonlinear processing properties of dendrites and structural plasticity (forming or breaking of synapses, re-routing of axonal branches etc.) in biological neurons. This solution was motivated by the fact that the liquid neurons in LSM produce sparse, high dimensional spike train outputs which was a key requirement in our earlier work on structural plasticity\cite{shaista_ijcnn1}. The method proposed in this paper, which we refer to as LSM with dendritically enhanced readout (LSM-DER), has the following benefits:
\begin{itemize}
	\item It can reduce the error as compared to \emph{p}-delta which is the current state of the art algorithm for the training of LSM readout\cite{Hour2011}, \cite{Vreeken2004}.
	\item It uses an order of magnitude less synaptic resources than parallel perceptrons while achieving comparable performance thus making it feasible for implementation in low-power smart sensors.
	\item The synapses connecting the liquid and readout can even have binary values without compromising performance. This is also very useful in hardware implementations since it removes the need for high resolution weights.
\end{itemize}

We have earlier presented a dendritic neuron with Network Rewiring (NRW) rule for classifying high dimensional binary spike rate patterns\cite{shaista_ijcnn1}. The primary differences in our current work compared to the earlier one are as follows:
\begin{itemize}
	\item We present a modified NRW rule that can be applied to arbitrary spike trains and not only rate encoded inputs.
	\item The modified rule can be used for solving approximation problems.
	\item For the first time, we demonstrate a dendritic neuron can be used as a readout for LSM.
	\item We demonstrate the stability of this architecture with respect to parameter variations making it suitable for low-power, analog VLSI implementations.
\end{itemize}

Some initial results for LSM-DER were presented earlier in \cite{roy_biocas1}. Here, we present a more detailed analysis of the reasons for improved performance of the algorithm as well as more results, a possible VLSI architecture and simulations to prove robustness of the proposed method to statistical variations plaguing VLSI implementations.

In the following section, we shall present a review about LSM, parallel perceptrons and the \emph{p}-delta learning rule. Then, we shall provide details about the non-linear neuron model and finally propose the Network Rewiring algorithm for LSM-DER. In Section III we will first discuss the classification and approximation tasks used to demonstrate the efficacy of the proposed architecture. Next, we shall provide the performance of LSM-DER on these problems and compare it with that of the traditional LSM-PPR. This section will also shed light on the dependance of the performance of LSM-DER on several key parameters. We will also present the robustness of the algorithm to variations in parameters, a quality that is essential for its adoption in low-power, sub-threshold neuromorphic designs which are plagued with mismatch. We will conclude the paper by discussing the implications of our work and future directions in the last section.

\section{Background and Theory}
In this section, we will first present, for the sake of completeness, some of the theory about the operation of the LSM, parallel perceptron readout and the \emph{p}-delta training algorithm. Then, we shall introduce our dendritically enhanced readout stage and the corresponding NRW algorithm for training it.
\subsection{Liquid State Machine}
LSM \cite{Maass2002} is a reservoir computing method developed from the viewpoint of computational neuroscience by Maass et al. It supports real time computations by employing a high dimensional heterogeneous dynamical system which is continuously perturbed by time varying inputs. The basic structure of LSM is shown in Fig.\ref{fig:LSM}. It comprises three parts: an input layer, a reservoir or liquid and a memoryless readout circuit.  The liquid is a recurrent interconnection of a large number of Leaky Integrate and Fire neurons (LIF) with biologically realistic parameters using dynamic synaptic connections in the reservoir. The readout is also implemented by a pool of LIF neurons which do not possess any interconnections within them. The LIF neurons of the liquid are connected to the neurons of the readout. The liquid does not create any output but it transforms the lower dimensional input stream to a higher dimensional internal state. These internal states act as an input to the memory less readout circuit which is responsible for producing the final output of the LSM.

Following \cite{Maass2002}, if $u(t)$ is the input to the reservoir then the liquid neuron circuit can be represented mathematically as a liquid filter $L^M$ which maps the input function $u(t)$ to the internal states $x^M(t)$ as:
\begin{equation}
\label{eq1}
x^M(t)=(L^Mu)(t)
\end{equation}

\begin{figure}
\begin{center}
\includegraphics[width=0.5\textwidth]{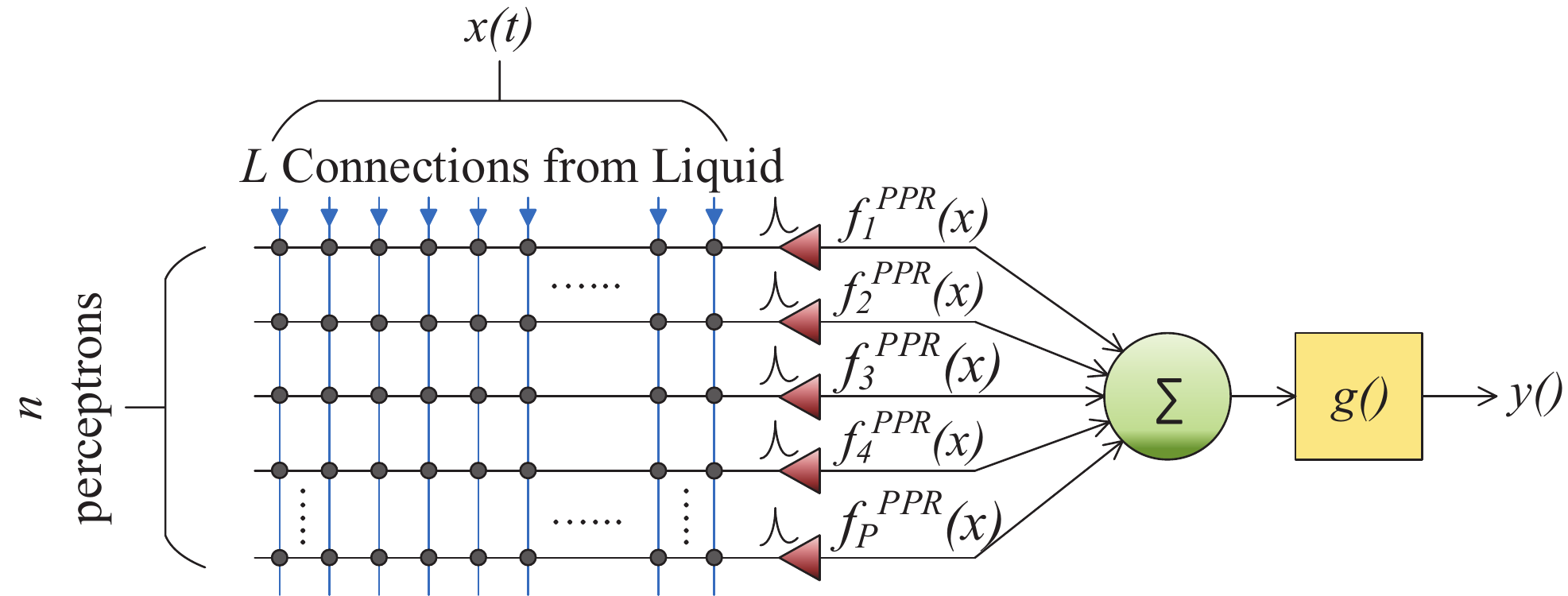}
\caption{In LSM-PPR, the readout stage is composed of a single layer of perceptrons which do not have any lateral connections. The output of the liquid is connected to each perceptron of this parallel perceptron stage.}
\label{fig:PPR}
\end{center}
\end{figure}
The next part of LSM i.e. the readout circuit takes these liquid states as input and transforms them at every time instant $t$ into the output $y(t)$ given by:
\begin{equation}
\label{eq2}
y(t)=f^M(x^M(t))
\end{equation}

For tasks with similar input activity level and time constant, the liquid is general whereas the readout is selected and trained in a task-specific manner. Moreover, multiple readouts can be used in parallel for extracting different features from the internal states produced by the liquid. For more details on the theory and applications of LSM, we invite the reader to refer to \cite{Maass2002}. Also, many research works have been published recently which focus on either the improvement of the LSM framework \cite{Xue2013,Ju2013,Hourdakis2013,Schliebs2012,Wojcik2012,Notley2012} or its applications in various real world problems \cite{Verstraeten,Goodman,Schliebs2012_cont,Probst2012,Jakimovski2011,elmir2011}. 

\subsection{Parallel Perceptron Readout and the p-delta learning algorithm}
\subsubsection{Parallel Perceptron Readout}
The readout stage employed in LSM-PPR is a layer of parallel perceptrons. A single layer composed of a finite number of perceptrons, each receiving the same input, is called a parallel perceptron as shown in Fig.\ref{fig:PPR}. A single perceptron with input \textbf{x}=$[x_1,$ $x_2..x_m]$, considering that the constant bias has been converted to one input, computes a function $f:\mathbb{R}^{m} \rightarrow \{-1,1\}$ defined as:
\begin{equation}
\label{pdelta}
f(\textbf{x})=
\begin{cases}
\; 1  	\text{ if $\textbf{w}.\textbf{x} \geq0$} \\
-1  \text{ otherwise}
\end{cases}
\end{equation}
where $\textbf{w} \in \mathbb{R}^{m}$ is the synaptic weight vector. To distinguish the neuronal functions in parallel perceptron readout (PPR) from dendritically enhanced readout (DER), we denote them by $f^{PPR}$ and $f^{DER}$ respectively. Let us now consider a parallel perceptron layer having $n$ number of perceptrons with outputs $f_1^{PPR}$, $f_2^{PPR}$,......,$f_n^{PPR}$ where $f_i^{PPR} :\mathbb{R}^{m} \rightarrow \{-1,1\}$. Then the output of the parallel perceptron readout is given by:
\begin{equation}
\widehat{o}=g(p)=g(\sum_{i=1}^n f_i^{PPR}(\textbf{x}))
\label{eq:ppr}
\end{equation}
where $p=\sum_{i=1}^n f_i^{PPR}(\textbf{x}) \in [-n,...,n]$ and $g:\mathbb{Z} \rightarrow \mathbb{R}$ is the squashing function. The function $g()$ is chosen according to the type of computation.

\subsubsection{The \emph{p}-delta learning algorithm}
PPR is trained by a simple yet efficient method termed as \emph{p}-delta rule proposed by Auer et al. in \cite{pdelta}. This simple rule can be utilized to approximate any boolean and continuous functions. Two advantages of this rule over traditional back propagation algorithm used in deep networks such as multi-layer perceptrons which can get stuck in local minima is that it is required to modify the weights of only a single layer of synapses and the computation and communication of high precision analog values are not required. We shall present some of the salient features of the algorithm here while inviting the reader to refer to \cite{pdelta} for details.

The \emph{p}-delta learning rule has two constituents, the first of which is the traditional delta rule that is employed to modify the weights of a `subset' of the individual neurons constituting the parallel perceptron layer. The next constituent is a rule which determines the subset. This rule states that the delta rule should be applied to those particular neurons which gives either the wrong output or the right output but with a small margin. Next, these two steps of the algorithm are discussed individually.

\paragraph{Obtaining Correct Outputs}

Let $\textbf{x}$, $\widehat{o}$ and $o$ be the input, output and the desired output respectively. Furthermore, let $\textbf{w}_1$,$\textbf{w}_2$,....,$\textbf{w}_n$ be the synaptic weight vectors of the $n$ perceptrons. If $\epsilon$ is the desired accuracy, then the output is considered correct if $|\widehat{o}-o|<\epsilon$. In this case, the weights need not be modified. On the other hand, if $\widehat{o}>o+\epsilon$, then the output is larger than expected. Thus, to reduce $\widehat{o}$, the number of weight vectors with $\textbf{w}_i.\textbf{x}\geq0$ are to be reduced. Application of the traditional delta rule to such a weight vector will give the update $\textbf{w}_i \leftarrow \textbf{w}_i + \eta \Delta _{i}$ where $\eta$ is the learning rate and $\Delta _{i}=-\textbf{x}$.

Proceeding similarly for the case where $\widehat{o}<o-\epsilon$, we can arrive at the general update rule $\textbf{w}_i \leftarrow \textbf{w}_i + \eta \Delta _{i}$ where $\Delta _{i}$ is given by:
 \begin{equation}
\Delta _{i}=
\begin{cases}
 -\textbf{x} \; if \; \widehat{o}>o+\epsilon \; \textrm{and} \; \textbf{w}_i.\textbf{x}\geq0 \\
 +\textbf{x} \; if \; \widehat{o}<o-\epsilon \; \textrm{and} \; \textbf{w}_i.\textbf{x}<0 \\
 \; 0 \:\: \textrm{otherwise}
\end{cases}
\end{equation}

\paragraph{Output stabilization}

In the preceding description, the weight vector of a particular perceptron is updated only when its output is incorrect. Hence, after the completion of the training phase, there are usually some weight vectors for which $\textbf{w}_i.\textbf{x}$ is very close to $0$. This implies a small perturbation of the input $\textbf{x}$ is capable of changing the sign of $\textbf{w}_i.\textbf{x}$ thereby reducing the generalization capabilities and the stability of the network output. In order to stabilize the output, the above rule needs to be modified in such a way that $\textbf{w}_i.\textbf{x}$ remains away from $0$. Thus, a new parameter $\gamma$ was introduced as the margin, i.e. the training tries to ensure $|\textbf{w}_i.\textbf{x}|>\gamma$.

The final learning rule described in \cite{pdelta} that incorporates all of these concepts is given by:
\begin{equation}
\label{eq:pdelta_margin}
\textbf{w}_i \leftarrow \textbf{w}_i + \eta
\begin{cases}
 (-\textbf{x}) \; if \; \widehat{o}>o+\epsilon \; \textrm{and} \; \textbf{w}_i.\textbf{x}\geq0 \\
 (+\textbf{x}) \; if \; \widehat{o}<o-\epsilon \; \textrm{and} \; \textbf{w}_i.\textbf{x}<0 \\
 \mu (+\textbf{x}) \; if \; \widehat{o}\leq o+\epsilon \; \textrm{and} \; 0\leq \textbf{w}_i.\textbf{x} < \gamma \\
 \mu (-\textbf{x}) \; if \; \widehat{o}\geq o-\epsilon \; \textrm{and} \; -\gamma < \textbf{w}_i.\textbf{x} < 0 \\
 \; 0 \:\: \textrm{otherwise}
\end{cases}
\end{equation}
\begin{equation}
\textbf{w}_i \leftarrow \textbf{w}_i/\|\textbf{w}_i\|
\end{equation}
where $\mu$ is a scaling factor. The parameters involved with \emph{p}-delta algorithm can be gradually modified with the progress of learning as presented in \cite{pdelta} and implemented in the LSM toolbox described in \cite{Natschlaeger02}.

\subsection{Model of Nonlinear Dendrites}\label{monondend}
\begin{figure}
\begin{center}
\includegraphics[width=0.5\textwidth]{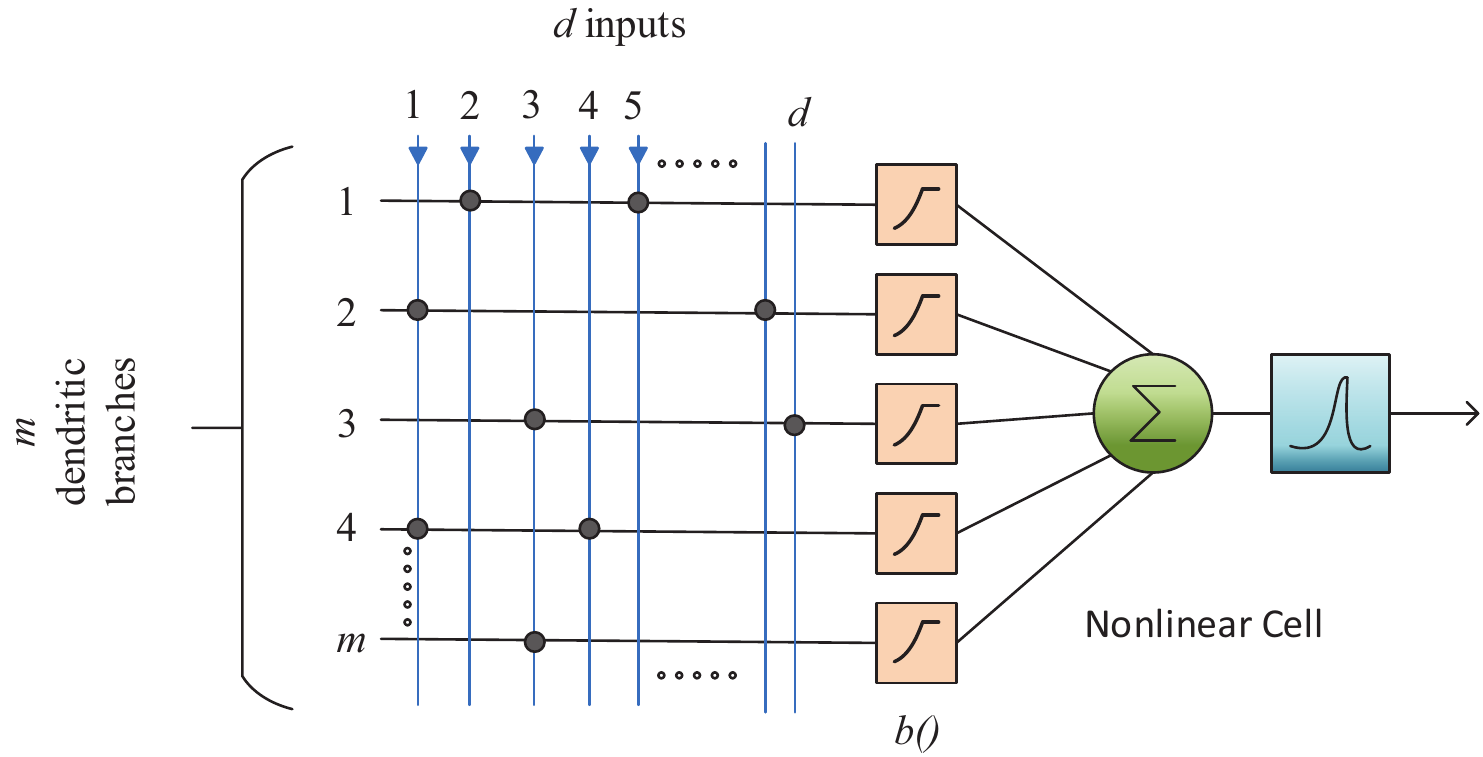}
\caption{A neuronal cell with active dendrites\label{fig:Model_NL}}
\end{center}
\end{figure}

Mel and Poirazi \cite{Mel2001} showed that neurons with active dendrites i.e. individual dendrites equipped with lumped nonlinearity possess higher storage capacity than their linear counterpart. Such a nonlinear neuronal cell (NL-cell), depicted in Fig.\ref{fig:Model_NL}, has $m$ identical branches connected to it with each branch having $k$ excitatory synapses. If \textbf{x} is an input vector to this system, then each synapse is excited by any one of the $d$ dimensions of the input vector where $d>>k$. The output response of $j^{th}$ dendritic branch is calculated as a nonlinear weighted sum of the currents of the $k$ synaptic points connected to the branch and is given by $z_j=b(\sum_{i=1}^k w_{ij} x_{ij})$, where $b()$ is the dendritic nonlinearity modeled as a nonlinear activation function, $w_{ij}$ is the synaptic weight of the $i^{th}$ synapse on the $j^{th}$ the branch and $x_{ij}$ the input arriving at that particular synaptic connection. We also define $v_j=\sum_{i=1}^k w_{ij} x_{ij}$ which is the weighted sum of the currents of the $k$ synaptic points of branch $j$ and is input to the branch's dendritic nonlinearity. Combining all the dendritic responses, the overall output $f(\textbf{x})$ of the neuronal cell is given by:

\begin{equation}
\label{eq:NLcell_eq}
 f(\textbf{x})=\sum_{j=1}^mz_j=\sum_{j=1}^mb(v_j)=\sum_{j=1}^mb(\sum_{i=1}^k w_{ij} x_{ij}).
\end{equation}
where $f()$ denotes the neuronal current-frequency conversion function.

LSMs are typically employed to solve two types of tasks: classification and regression. For both these tasks, we employ two NL-cells and calculate the output by noting the difference of the output of the two NL-cells. The overall output of the circuit is given by
\begin{equation}
\label{eq3}
y=g[f_+(\textbf{x})-f_-(\textbf{x})]
\end{equation}

\subsection{Liquid State Machine with dendritically enhanced readout}
\label{lsm_der}

\begin{figure}
\begin{center}
\includegraphics[width=0.5\textwidth]{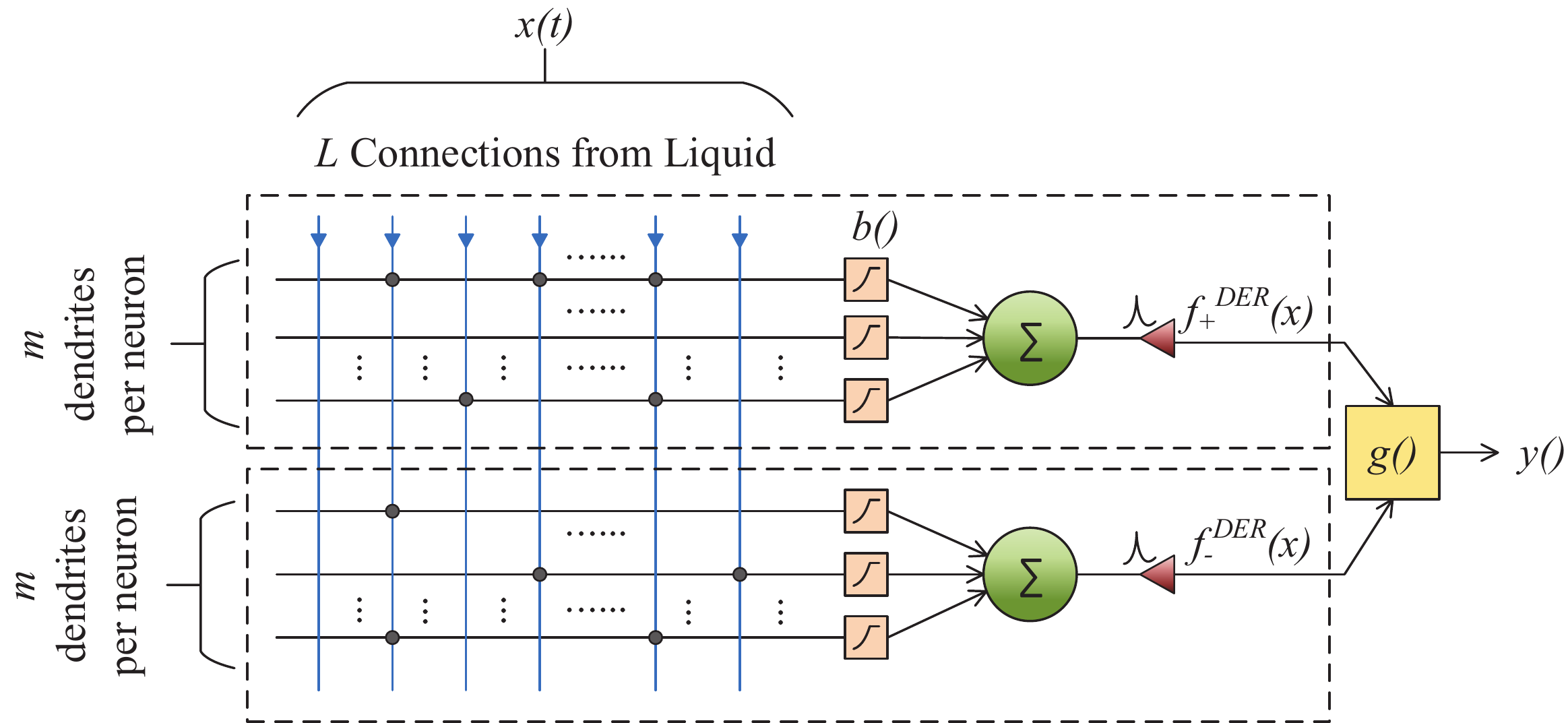}
\caption{In LSM-DER, the readout stage following the liquid is composed of two NL-cells. The final readout output y() is obtained by taking the difference of the output of the two cells and passing it through the function g().}
\label{fig:DER}
\end{center}
\end{figure}

In LSM-DER, the liquid described in \cite{Maass2002} is followed by the two neuronal cell architecture depicted in Fig.\ref{fig:DER}. Here, we denote the output of each cell as $f_{+/-}^{DER}$ with the added superscript DER to contrast with the earlier figure for LSM-PPR. The earlier readout circuit consisting of the parallel perceptrons has now been replaced with the above mentioned circuit. 

For training the two-cell architecture, a much simpler and hardware friendly version of the learning algorithm proposed in \cite{Mel2001} has been employed. The key point of the NRW learning algorithm is removing a synapse which is contributing most to the classification errors and replacing it with a new synapse formed from a different afferent line. During training the NL-cells, a global teacher signal $t$ is also presented indicating the desired output to be obtained on the application of a particular input sample. If error $e=(t-y)$, then according to the gradient-descent algorithm the weight modification, $\triangle w_{ij}$ is given by
\begin{equation}
\begin{aligned}
\triangle w_{ij}= & -\frac{\partial e^2}{\partial w_{ij}}\\
                = & 2<(t-y) \frac{\partial y}{\partial w_{ij}}> \\
                = & 2<(t-y) \frac{\partial g(f_+(x)-f_-(x))}{\partial w_{ij}}>
\end{aligned}
\end{equation}
where $<.>$ signifies averaging over the entire training set. Hence, the weight modification for the positive and the negative neuronal cells are $\triangle w_{ij}=2<(t-y)g'b_j'x_{ij}>$ and $\triangle w_{ij}=-2<(t-y)g'b_j'x_{ij}>$  respectively where $g'$ and $b'$ are the derivatives of $g()$ and $b()$ respectively. However, we have used binary synapses in this work for robust hardware implementation. This implies our NRW learning procedure does not need any weight modification but requires the formation and elimination of synapses. Hence, we consider a fitness parameter $\phi_{ij}=\triangle w_{ij}$ to guide this process. The $\phi_{ij}$ used in \cite{Mel2001} was given by $\phi_{ij}=<x_{ij}b_j'g'signum(t-y)>$.Thus, an existing synapse with a low value of $\phi_{ij}$ needs to be eliminated. The learning procedure dictates the replacement of this poorly performing synapse with a synapse having a high value of $\phi_{ij}$ from a randomly chosen replacement set. The NRW learning rule thus creates a morphological change of the dendrites guided by the gradient descent rule. From the viewpoint of hardware implementation, we have further simplified the learning rule to use a performance index:

 \begin{equation}
 \label{eq4}
 c_{ij}=
 \begin{cases}
 <x_{ij}b_{j}'(t-y)> \;\;\;\;\;\;\textrm{for positive cell} \\
 -<x_{ij}b_{j}'(t-y)>\;\; \textrm{for negative cell}
 \end{cases}
 \end{equation}

This is simpler since it does not require computing derivatives of $g()$ as in \cite{Mel2001}. The second difference from \cite{Mel2001} is the particular functional form of dendritic nonlinearity $b()$ that we used to simplify $c_{ij}$ further. The dendritic nonlinearity $b()$ used in our simulations was $b(x) = x^2/x_{thr}$ implying $b'(x)=2x/x_{thr}$. According to the notation for our model in \ref{eq:NLcell_eq}, $b'(v_j)=2v_j/x_{thr}$. Hence, by ignoring the constants we finally get $c_{ij}=<x_{ij}v_j(t-y)>$ for the positive cell. This function can be easily implemented on-chip without needing extra calculations to obtain derivative of $b()$ since $v_j$ is already being computed for normal operation. Though we have not implemented spike based learning in this work, the motivation of choosing this learning rule is to utilize circuits that can compute these correlations \cite{Brink2012} on-chip in future.

However, a square law output of each dendritic branch will result in unrealistically large values for large inputs. In earlier work \cite{Mel2001}, the authors had used nonlinearities like $b(x)=x^{10}$ which also have the same issue and will definitely be a problem in VLSI implementations, either analog or digital. Hence, we included a saturation level, $x_{sat}$ at the output such that for $b(x) > x_{sat}$, $b(x) = x_{sat}$. This models a more realistic scenario in hardware and also leads to power saving in analog hardware if the nonlinearity $b()$ is implemented in current mode. Incidentally, biological neurons also exhibit a saturating nonlinearity\cite{polsky_mel_04} due to similar constraints.

The output of the liquid were applied as input pattern to the setup of Fig.\ref{fig:DER}. Thus, the dimension of the input pattern is now equal to the number of liquid neurons. For each $m$ branches, $k$ synaptic contacts with weight $1$ were formed by randomly selecting afferents from one of the $d = L$ input lines, where $L$ is the number of liquid neurons. The learning process comprised the following steps in every iteration:

\begin{enumerate}
	\item For each pattern of the entire training set, the outputs of both the neuronal cells i.e $f_{+}^{DER}$ and $f_{-}^{DER}$ were calculated. This was followed by computing the overall response of the classifier as per Equation \ref{eq3}.
	\item After each application of the entire training set, the Mean Absolute Error (MAE) was calculated as $\sum|(t-y)|/P$, where P is the size of the training set.
	\item A random set $T$ of $n_T$ synapses were selected from $k\times m$ existing synapses per cell for probable replacement. The performance index $c_{ij}$ corresponding to $i^{th}$ synapse of the $j^{th}$ branch was calculated for each synapse in the set $n_T$ for both the neuronal cells.
	\item The synapse with the lowest value of performance index ($c_{ij}$) in $T$ was labeled for replacement with the synapse with the highest value of performance index from an another randomly chosen replacement synapse set $R$ having $n_R$ of the $d$ input lines. The set $R$ was created by placing $n_R$ `silent' synapses from $d$ input lines on the branch with the lowest $c_{ij}$ synapse. They do not contribute to the calculation in step (1).
	\item The synaptic connections were updated if the replacement led to a decrease in MAE. If there are no such reduction of MAE, a new replacement set $R$ is chosen. If $max_{loc}$ such choices of $R$ do not reduce MAE, it is assumed that the algorithm has stuck to a local minima and connection changes are made in an attempt to escape the local minima even if it increases the MAE .
	\item The above mentioned steps were repeated for a $max_{iter}$ number of iterations after which the algorithm is terminated and the connection corresponding to the best minima among all the iterations is saved as the final connection.
	
\end{enumerate}

\begin{figure}
\begin{center}
\includegraphics[width=0.5\textwidth]{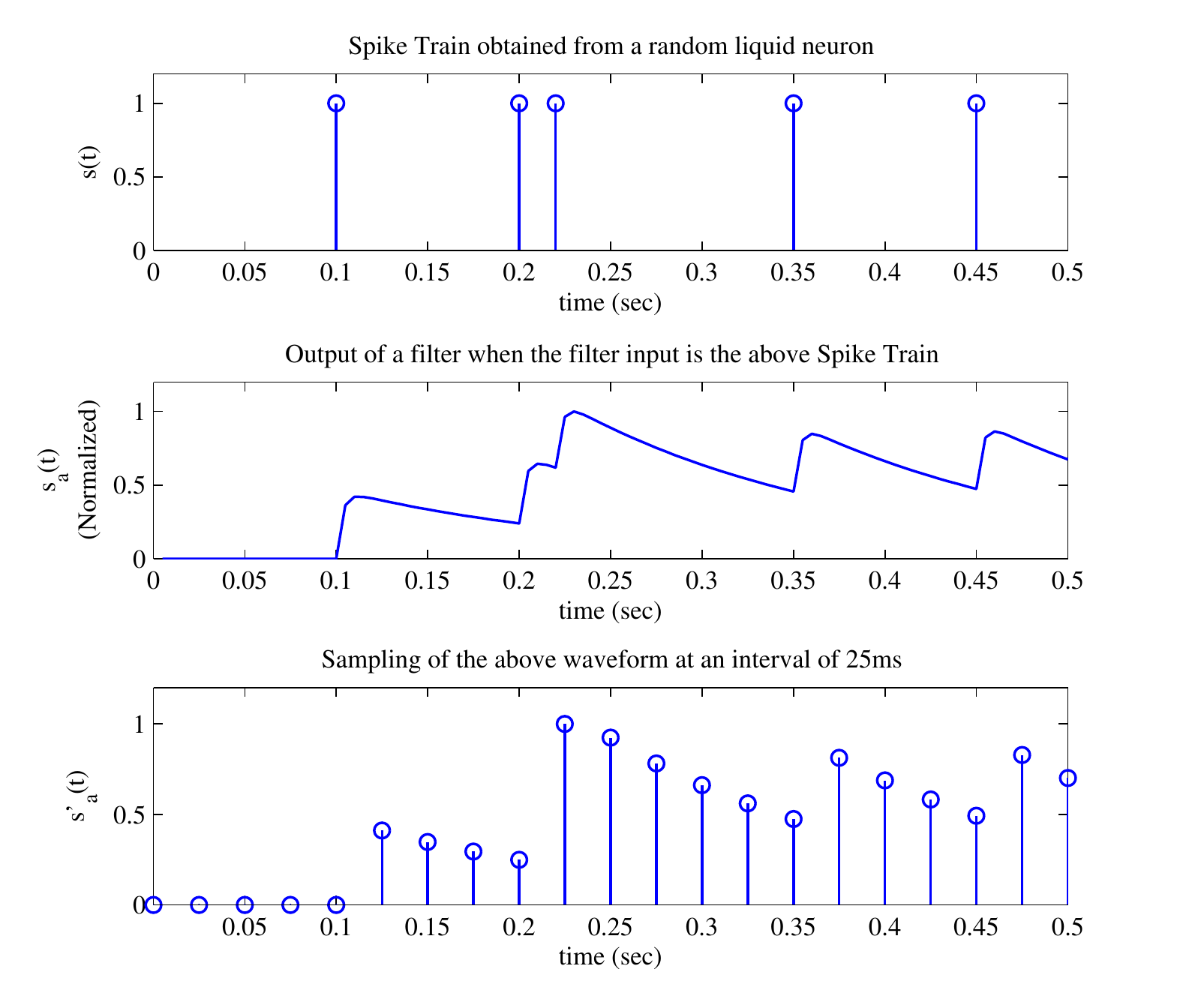}
\caption{An example of input generation for the proposed readout where the spike train $s(t)$ is first convolved with a kernel representing post-synaptic current to produce $s_a(t)$ and then sampled at a desired temporal resolution to give $s_a'(t)$.}
\label{fig:inpgen}
\end{center}
\end{figure}

\begin{figure}
\begin{center}
\includegraphics[width=0.5\textwidth]{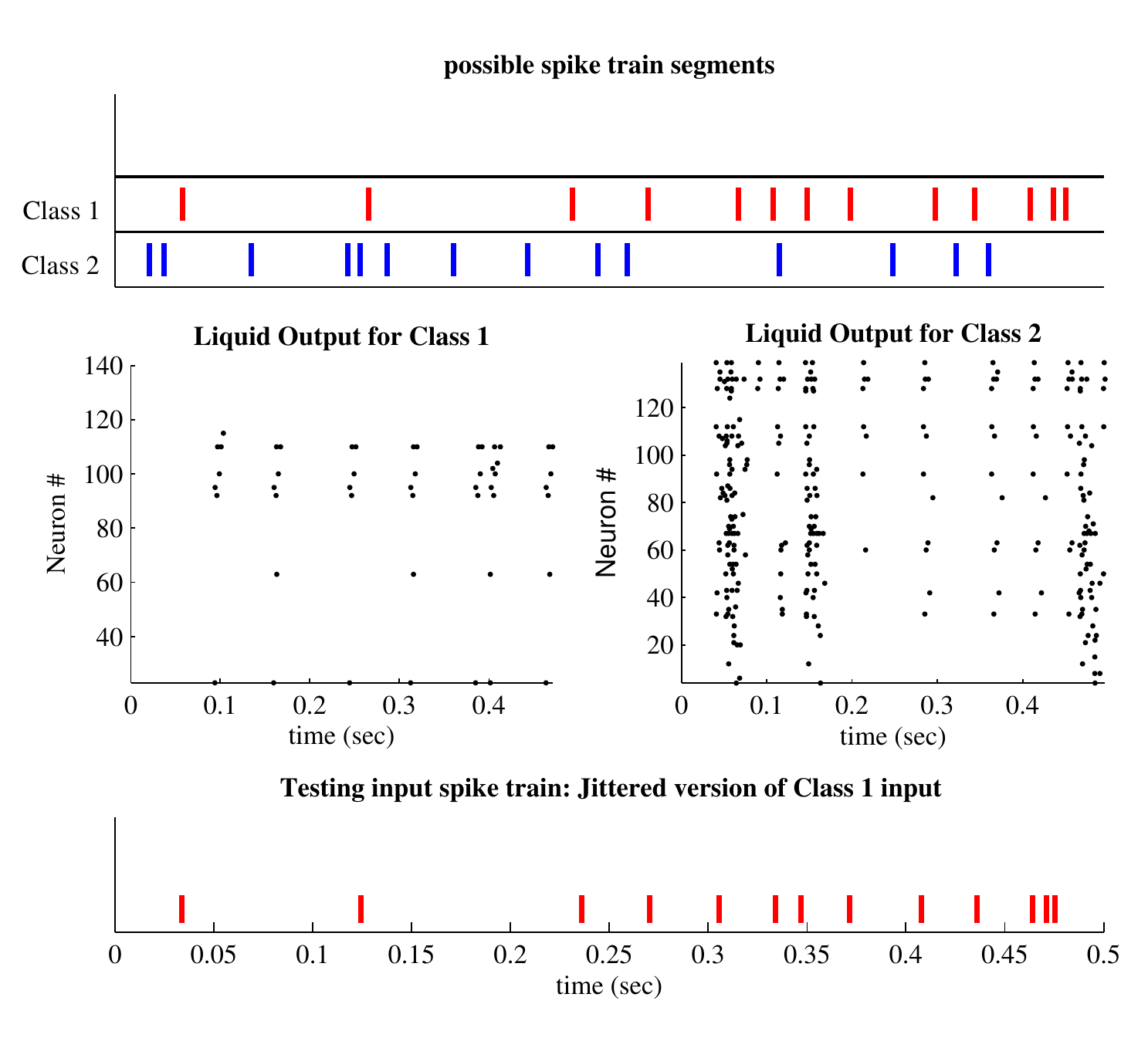}
\caption{Task I is the classification of spike trains. Two different classes of spike trains and the corresponding liquid output when these spike trains are projected into it are shown. Also, a jittered version of a spike train belonging to Class 1 that will be used for testing is depicted.}
\label{fig:spikeclass}
\end{center}
\end{figure}

\begin{figure}
\begin{center}
\includegraphics[width=0.5\textwidth]{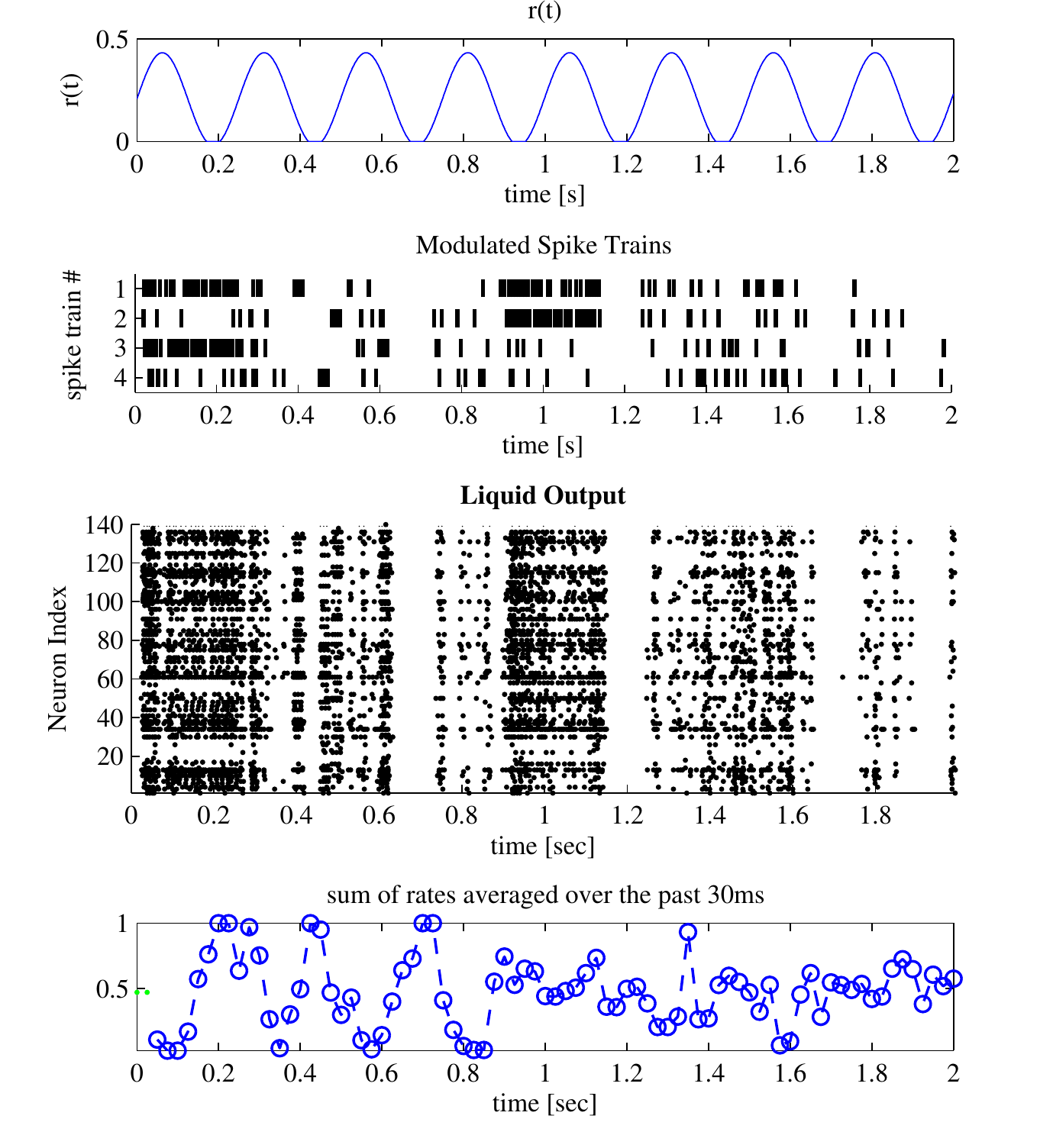}
\caption{Task II involves the approximation of a desired function. r(t) is a signal that modulates $4$ Poisson spike trains to generate the liquid input. The modulated spike trains to be given as input to the liquid and  the liquid output are shown. The sum of rates corresponding to the modulated spike trains is also shown\label{fig:sumrates}}
\end{center}
\end{figure}

\subsection{Spike trains as input to Neurons with active dendrites}
\label{sec:spike_gen}
The output of the liquid neurons are spike trains, while the earlier theoretical description of the NL cells had inputs $\textbf{x} \in \mathbb{R}^{d}$. Thus, to use the NL cells to act as readout, we need a method to transform spike trains to act as inputs to the NL cells. This process is described next.

Suppose, the arbitrary spike train $s(t)$ is given by:
\begin{equation}
s(t)=\sum_{t_f} \delta(t-t_f)
\end{equation}
where $t_f$ indicate the spike firing times. One way to convert this to an analog waveform is to convolve it with a low-pass filtering kernel. We choose a fast rising, slow decaying kernel function, $h(t)$, that mimics post-synaptic current (PSC) waveform and is popularly used in computational neuroscience\cite{tempotron}. The specific form of the function we use is given by:
\begin{equation}
h(t)=I_0(e^{-\frac{t}{\tau_s}}-e^{-\frac{t}{\tau_f}})
\end{equation}
where $\tau_f$ and $\tau_s$ denote the fast and slow time constants dictating the rise and fall times respectively and $I_0$ is a normalizing factor. Hence, the final filtered analog waveform, $s_a(t)$ corresponding to the spike train s(t) is given by:
\begin{equation}
s_a(t)=\sum_{t_f} h(t-t_f)
\end{equation}
Finally, to train the NL-cells in the readout, we need a set of discrete numbers which we obtain by sampling $s_a(t)$ at a desired temporal resolution $T_s$. The sampled waveform, $s'_a(t)$ is given by:
 \begin{equation}
s'_a(t)=s_a(t)\sum_i \delta(t-iT_s)
\end{equation}
Therefore, for the $L$ spike trains produced by the liquid neurons, if the temporal duration of the spike trains are $T$, then it will result in a total of $\lfloor T\setminus T_s \rfloor$ samples $\bf{x_i} \epsilon \mathbb{R}^{L}$. For our simulations, we have chosen the temporal resolution $T_s=25$ ms. The whole process is shown for a spike train output from one random liquid neuron in Fig.\ref{fig:inpgen}.

\section{Experiments and Results}
\subsection{Problem Description}
In this sub-section, we describe the two tasks used to demonstrate the performance of our algorithm. The reason for this choice is that both of these are standard problems that are shown in the original publication on LSM \cite{Maass2002} and are included as examples in the LSM toolbox \cite{Natschlaeger02}. Also, we chose one task to be a classification and the other to be an approximation since they are representative of the class of problems solved by LSM.
\subsubsection{Task I: Classification of spike trains}
The first benchmark task we have considered is the Spike Train Classification problem \cite{Natschlaeger02}. The generalized Spike Train Classification problem includes $q$ arrays of $e$ Poisson spike trains having frequency $f$ and length $T_{max}$ which are labeled as templates $1$ to $q$. These spike trains are used as input to the LSM, and the readout is trained to identify each class. Next, a jittered version of each template is generated by altering each spike within the template by a random amount that is stochastically drawn from a Gaussian distribution with zero mean and standard deviation $STD$. This $STD$ is termed as the $jitter$. Given a jittered version of a particular spike train, the task is to correctly identify the class from which it has been drawn. In this article, we have considered $q=2$, $e=1$, $f=20$, $STD=4$ ms and $T_{max}=0.5$ sec. Fig.\ref{fig:spikeclass} shows an example instance of two classes of spike trains and the output of the liquid when jittered versions of these spike trains are injected into it. The figure also shows a jittered input spike train to be used for testing. The training and testing sets are composed of same number of patterns.

\subsubsection{Task II: Retrieval of sum of rates}
The next task is more difficult than the previous classification problem. The job of the network is to produce at its output the sum of firing rates of the input spike trains averaged over a past time window. $e$ Poisson spike trains are injected into the liquid, the firing rates of which are modulated by a randomly chosen function $r(t)=A+B sin(2 \pi f t + \alpha)$ lying in the range $(0,1)$. The parameters $A$, $B$ and $f$ were drawn randomly from the following intervals: $A$ [0 Hz, 30 Hz] and [70 Hz, 100 Hz], $B$ [0 Hz, 30Hz] and [70 Hz, 100 Hz], $f$ [0.5 Hz, 1 Hz] and [3 Hz, 5 Hz]. The phase was fixed at $\alpha = 0 $ deg. To generate the test input by a different distribution than the training examples the values of $A$, $B$ and $f$ in the testing case were kept as $50 Hz$, $50 Hz$ and $2 Hz$ respectively. The values of these parameters for the testing case were chosen in such a way that they lie in the middle of the gaps between the two intervals used for these parameters during training. At any point of time $t,$ the job of the network is then to give as output the normalized sum of rates averaged over the interval $(t-D-W,t-D)$ where the width of the interval is $W$ and $D$ is the delay. In our case we have taken $e=4$, $W= 30 ms$ and $D=0$ i.e. no delay. Fig.\ref{fig:sumrates} shows the $4$ input Poisson spike trains, the function $r(t)$, the modulated spike trains and the liquid output when the modulated spike trains are projected into the liquid. The bottom plot in Fig.\ref{fig:sumrates} shows the target function i.e. the sum of rates averaged over the last $30$ms.

\subsection{Choice of Parameters}

The values of the parameters used by the LSM-DER architecture and NRW learning rule are reported in Table \ref{table:params}. We shall next discuss the procedure for selecting these parameters.

\paragraph{Total number of synapses per neuronal cell (s)}
The number of synapses required for connecting the liquid to the readout in case of LSM-PPR is $L \times n$ where $L$ and $n$ are the number of liquid neurons and number of perceptrons in the readout stage respectively.  To demonstrate that LSM-DER uses synaptic resources efficiently, we employ for LSM-DER the same number of synaptic resources used by LSM-PPR when $n=1$ i.e. $L$ number of synapses.  As described earlier, DER comprises two neuronal cells (to eliminate the need for negative weights); thus each has half of the total synapses i.e. $\frac{L}{2}$ synapses are allocated for the positive cell and $\frac{L}{2}$ for the negative cell.

\paragraph{Number of dendrites per neuronal cell (m)}
In \cite{Mel2001} a measure of the pattern memorization capacity,$B_{N}$, of the NL-cell (Fig.\ref{fig:Model_NL}) has been provided by counting all possible functions realizable as:

\begin{equation}
\label{eq:BN}
B_{N} = log_2 \binom{\binom{k+d-1}{k}+m-1}{m} bits
\end{equation}

where $s$, $m$, $k$ and $d$ are the total number of synapses, the number of dendrites, the number of synapses per dendrites and the dimension of the input respectively for this neuronal cell. As the readout of LSM-DER employs two such opponent cells, thus the overall capacity is twice this value.

\begin{figure}
\begin{center}
\includegraphics[width=0.5\textwidth]{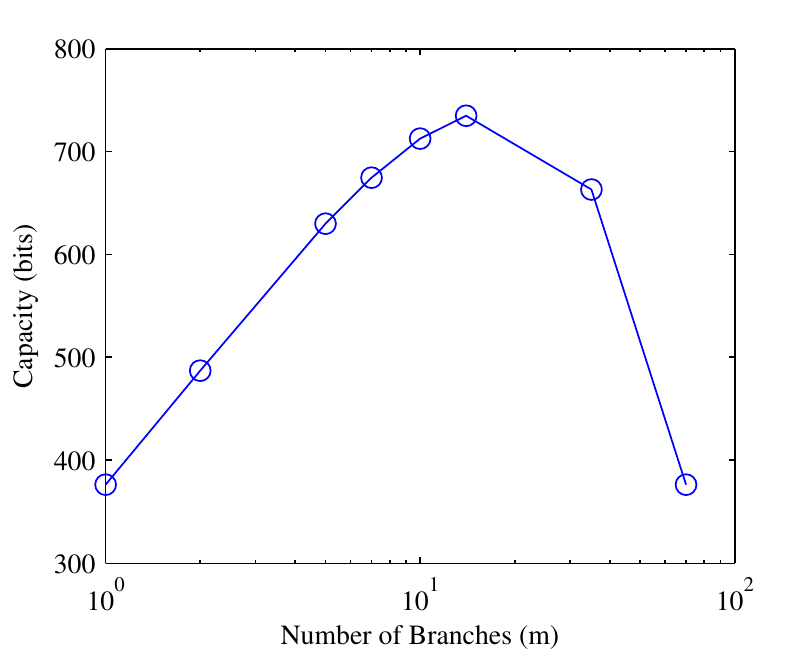}
\caption{Pattern memorization capacity of a NL-cell ($B_N$) is plotted as function of the number of dendritic branches ($m$) for a fixed number of synapses ($s$).\label{fig:capacity}}
\end{center}
\end{figure}

In our case $d = L$ and $s = \frac{L}{2}$. Since $s=m \times k$, for a fixed $s$ all possible values which $m$ can take are factors of $s$. We calculate $B_N$ for these values of $m$ by Equation \ref{eq:BN} and show it in Fig.\ref{fig:capacity}. It is evident from the curve that the capacity is maximum when $m=14$. But, in our simulations, we found that the learning algorithm cannot train the NL-cell to attain this maximum capacity in a reasonable time due to the huge number of possible wiring configurations to choose from. Hence, as a compromise between capacity and trainability, we chose $m=7$ in the following simulations.

\paragraph{Number of synapses per branch (k)}
After $s$ and $m$ has been set, the value of $k$ can be calculated as $k=\frac{s}{m}$.

\paragraph{The slow ($\tau_s$) and fast time constant ($\tau_f$)}
The fast time constant ($\tau_f$) and the slow time constant $\tau_s$ have been defined in Section \ref{sec:spike_gen}. $\tau_f$ usually takes a small value in hardware realizations and is not tuned. As for  $\tau_s$, if its value is too small, then the post synaptic current due to individual spikes die down rapidly and thus temporal summation of separated inputs do not take place. On the other hand large values of $\tau_s$ renders all spikes effectively simultaneous. So, in both extremes the extraction of temporal features from the liquid output is impaired. A detailed discussion on the selection of $\tau_s$ with respect to the inter spike interval (ISI) of the liquid output is given in Section \ref{dis}.

\paragraph{Threshold of nonlinearity: $x_{thr}$}
It can be seen that the value of $x_{thr}$ (defined in Section \ref{lsm_der}) is different for the two tasks. For selecting $x_{thr}$, we need to note that the operating principle of the NRW learning rule is to favor those connection topologies where correlated inputs for synaptic connections on the same branch. The nonlinear function $b()$ should give a supra linear output when more than one synaptic inputs on the same branch are co-activated. For the nonlinear function $b(x)=\frac{x^2}{x_{thr}}$used here, $b(x)>x$ for $x>x_{thr}$. Hence, the choice of $x_{thr}$ is given by:
\begin{equation}
\label{eq:xthr}
\overline{I_{syn}}<x_{thr}<2\overline{I_{syn}}
\end{equation}
where $\overline{I_{syn}}$ denotes the average post-synaptic current from an active synapses. Performing this calculation for a large number of input patterns, we obtained the values of $\overline{I_{syn}}=1.65$ for Task I and $\overline{I_{syn}}=5.34$ for Task II. Thus, in our case we chose the value of $x_{thr}$ as $1.8$ for Task I and $7$ for Task II.

Moreover, an extensive study on the performance of the algorithm due to the variation of $x_{sat}$ will also be provided. For fair comparison, the number of liquid neurons used for both LSM-DER and LSM-PPR was $140$. In case of LSM-PPR, the readout stage consists of $n=40$ neurons which implies the use of $140\times 40$ synapses to connect the liquid neurons to the array of parallel perceptrons. Unless otherwise mentioned, these are the default values of parameters in this article.
\begin{table}[ht]
\caption{Parameter Values}
\centering
\begin{tabular}{|c|p{4cm}|c|c|}
  \hline
  Parameters & Description & Task I & Task II\\ \hline
  $m$ & Dendrites per neuronal cell & 7 & 7\\
  $k$ & Synapses per dendrite & 10 & 10\\
  $L$ & Number of Liquid neurons & 140 & 140\\
  $P$ & Number of patterns in training set & 200 & 200 \\
  $\tau_s$ & Slow time constant & 7.5ms & 7.5ms\\
  $\tau_f$ & Fast time constant & 30ms & 30ms\\
  $n_T$ & Number of synapses in the target set for probable replacement & 15 & 15\\
  $n_R$ & Number of synapses in the replacement set & 25 & 25\\
  $max_{iter}$ & Maximum number of iterations & 1000 & 1000\\
  $max_{loc}$ & Maximum number of iterations required to declare local minima & 30 & 30\\
  $x_{thr}$ & Threshold of nonlinearity & 1.8 & 7\\
  $x_{sat}$ & Dendritic branch saturation Level & 75 & 75\\

  \hline
\end{tabular}
\label{table:params}
\end{table}

\paragraph{Choice of g()}
For Task I, one of the neuronal cells was trained to respond to $'+'$ patterns and the other to $'-'$ patterns. Thus, $g()$ in this case is $signum():\mathbb{R} \rightarrow \{0,1\}$ which operates on the combined activity of the two neurons to decide the category ($+$ or $-$) of the input pattern. The output of the $signum$ function i.e. $y$ can take a value of either $1$ or $0$ implying that the pattern belongs to $'+'$ or $'-'$ category respectively. For Task II, $g()$ is given by:
\begin{equation}
\label{eqgx}
g(z)=\frac{1}{1+exp(-(z/2))}
\end{equation}

\subsection{Results: Performance of LSM-DER and NRW algorithm}
\label{results}
\begin{figure}
\begin{center}
\includegraphics[width=0.45\textwidth,height=4cm]{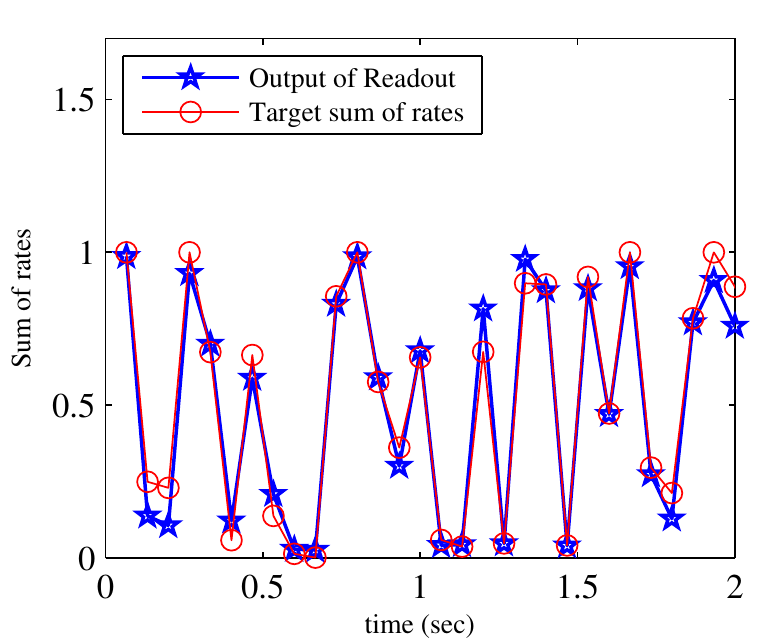}
\caption{The output produced by the LSM-DER after training is superimposed on the target function for a randomly chosen time window and show a very close match.}
\label{fig:comp_tar_out_sor}
\end{center}
\end{figure}

The proposed readout is separately trained for Task I and Task II. Similar to the method followed in \cite{Natschlaeger02,Maass2002}, we calculate mean absolute error (MAE) by averaging the error in approximation or classification across all the patterns. We first demonstrate the convergence of the NRW algorithm by plotting in Fig.\ref{fig:comp_tar_out_sor} both the target function and the readout output after training (during a randomly selected time window) of LSM-DER for a test pattern in Task II. The figure shows that the readout is able to approximate the desired sum of rates very closely after convergence. Instead of the $(t-y)$ term if we use $signum(t-y)$ like \cite{Mel2001,shaista_ijcnn1}in $c_{ij}$ then the MAE obtained by LSM-DER for Task II is 0.1496 instead of 0.0923.

\begin{figure*}[t]
	\centering
	\subfloat[]{\includegraphics[width=0.4\textwidth,height=5cm]{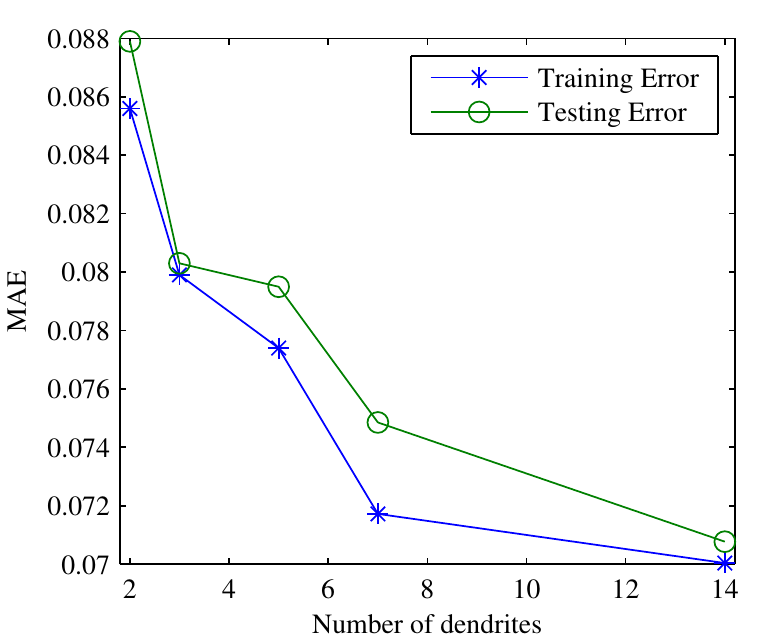}}\qquad
	\subfloat[]{\includegraphics[width=0.4\textwidth,height=5 cm]{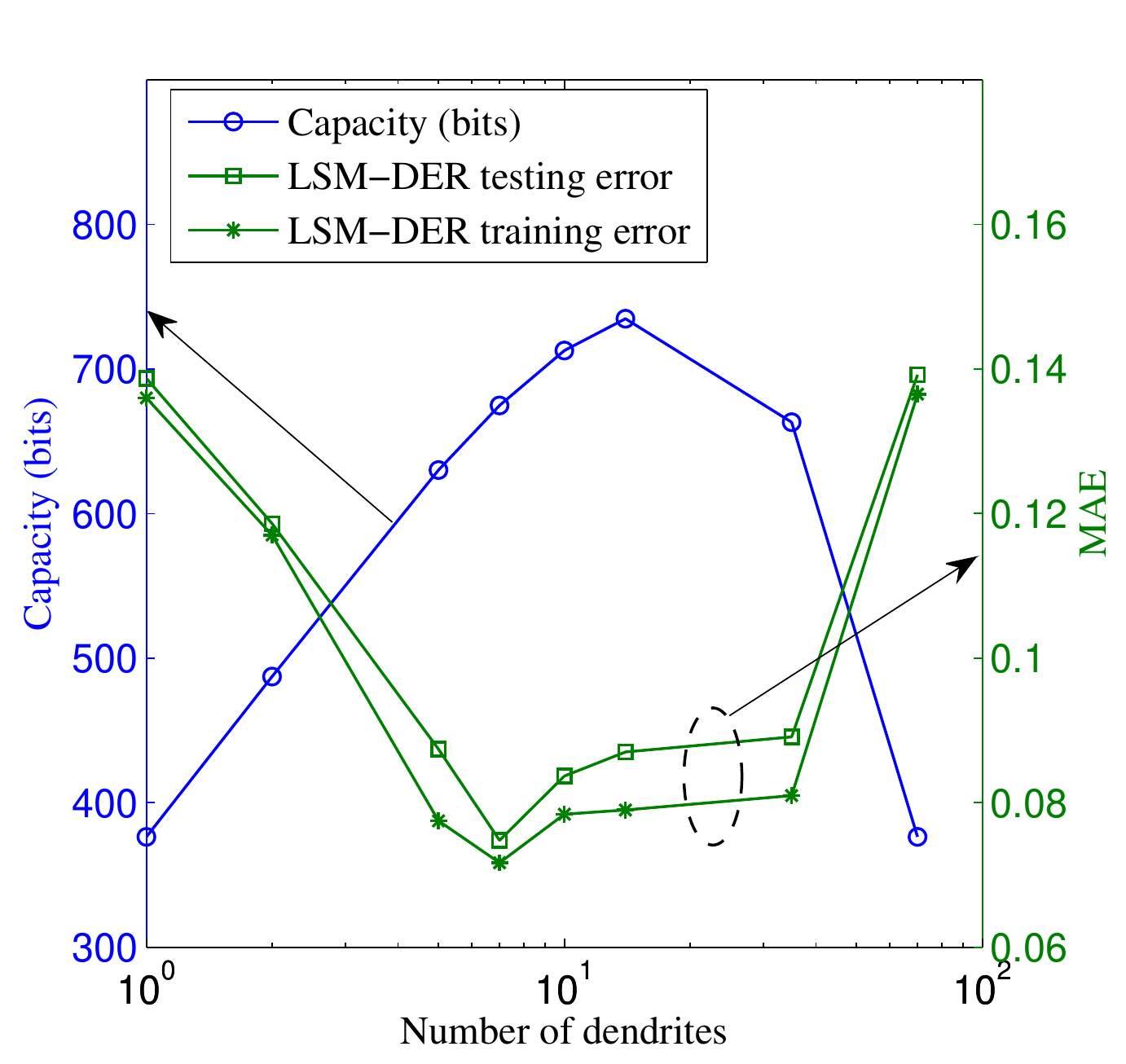}}\\
	\subfloat[]{\includegraphics[width=0.4\textwidth,height=5cm]{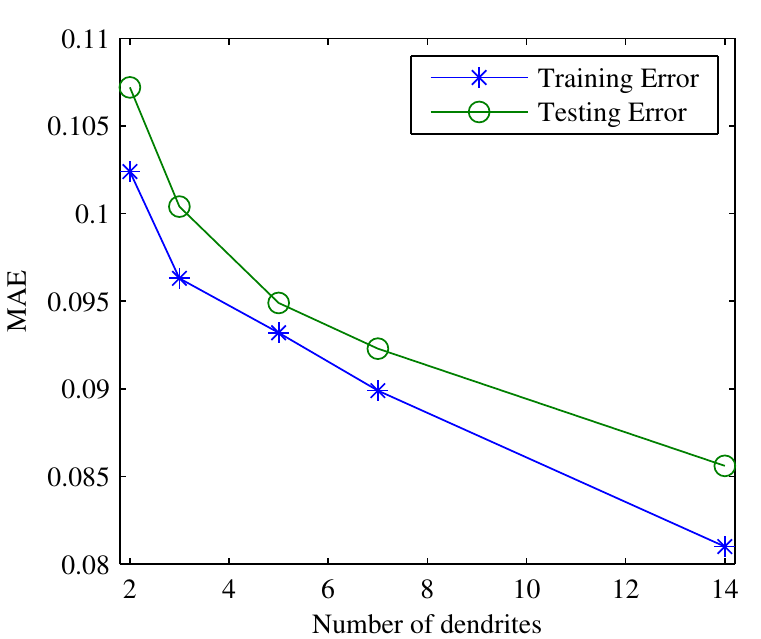}}\qquad
	\subfloat[]{\includegraphics[width=0.4\textwidth,height=5cm]{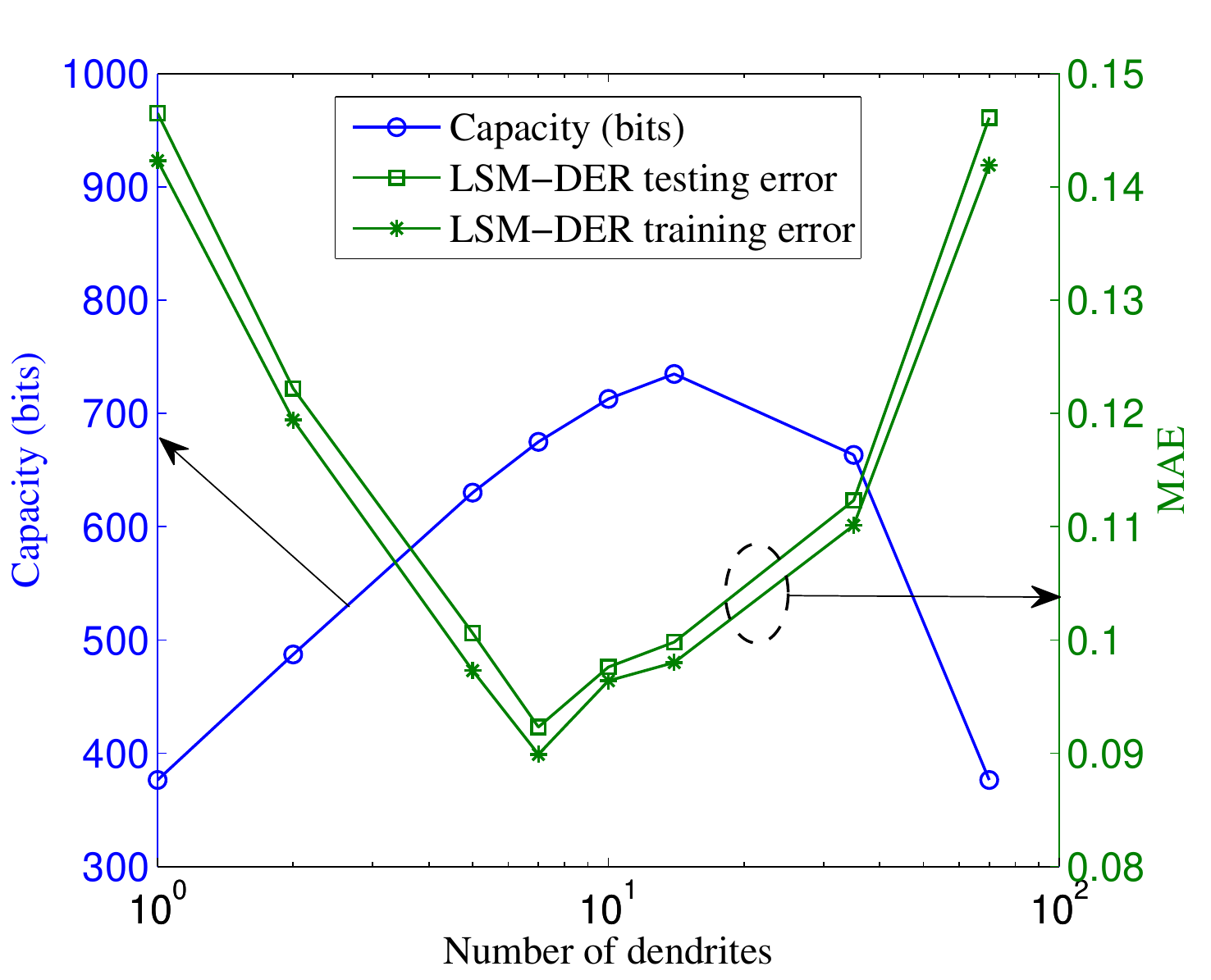}}
	\caption{Keeping the number of synapses per dendritic branch ($k$) of the neuronal cells constant at $10$, the average MAE over 10 trials is plotted for (a) Task I and (c) Task II when the number of dendrites ($m$) is varied. Thus, the total number of synapses connecting the reservoir with the readout increases with $m$ and results in a reduction of error. When the total number of synapses for each neuronal cell ($s/2$) is kept constant at $70$, MAE is plotted for Task I (b) and Task I (d) when number of dendrites are increased. In this case, the error initially reduces, reaches a minima and then starts increasing.}\label{fig:msedend}
\end{figure*}

Similar to \cite{Mel2001}, it is important to understand the variation in error for both tasks as the architecture of the LSM-DER (characterized by $m$ and $k$) is varied. We have performed two different experiments to study this dependence. In the first experiment, the number of dendritic branches ($m$) is varied while keeping the number of synaptic connections per branch ($k$) at a constant value of $10$ (other parameters are fixed at the values mentioned in table \ref{table:params}). This results in an increase in the total number of synaptic connections $s=m\times k$ from the liquid to each readout neuron as $m$ is increased. The results for this experiment are shown in Fig. \ref{fig:msedend}(a) and (c) for tasks I and II respectively. As expected, the MAE reduces with increasing $m$ since an increase in $m$ with constant $k$ results in more possible functions.

In the second experiment, we vary $m$ while keeping the total number of synapses allocated to a neuronal cell, $s$ constant at a value of $70$. The results for this procedure are plotted in Fig. \ref{fig:msedend} (b) and (d) for tasks I and II respectively. The capacity of the readout of LSM-DER ($B_N$) is also plotted in these figures. The figures show that as $m$ increases $B_N$ first increases and attains a maxima after which $B_N$ decreases. The MAE should have been the lowest when $B_N$ attains the maximum value i.e. at $m=14$. But, this does not happen in practice and MAE is lowest for $m=7$. We suspect this is because at $m=10$ and $m=14$, the total number of distinct input-output functions that can be implemented by rewiring becomes too large and the NRW algorithm easily gets trapped in several equivalent local minima. We are currently trying to develop better optimization strategies to overcome this issue.

\begin{figure}[!t]
\centering
   \subfloat[]{\includegraphics[width=0.25\textwidth]{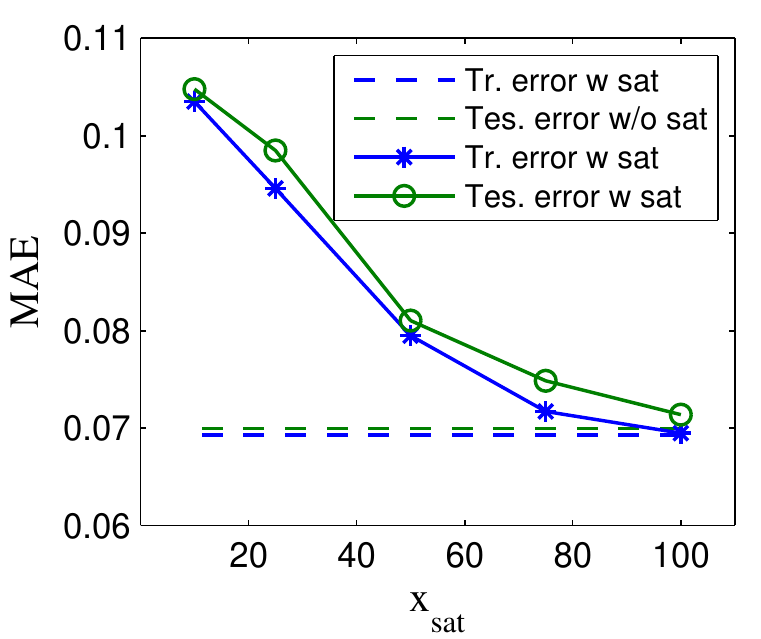}}
   \subfloat[]{\includegraphics[width=0.25\textwidth]{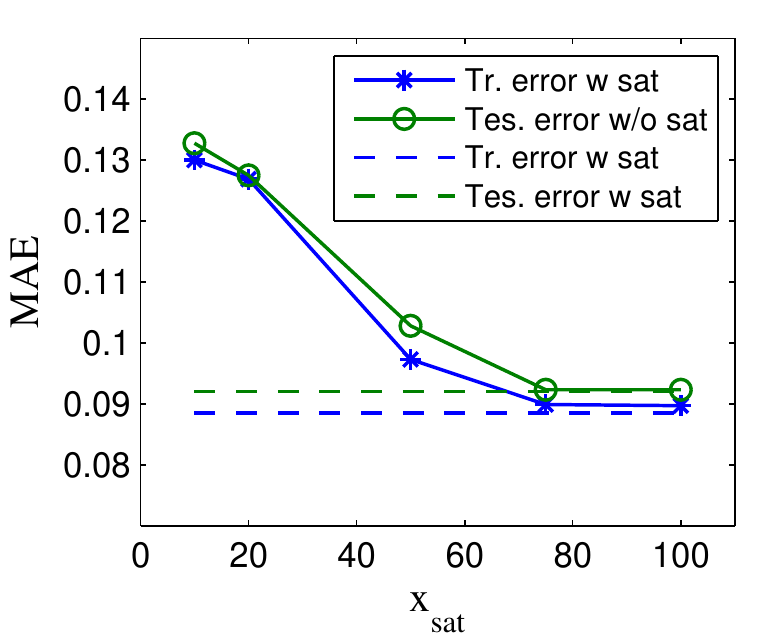}}
\caption{Change in MAE, averaged over 10 trials, when $x_{sat}$ is varied for (a) task I and (b) task II. The dashed lines indicate the MAE for $x_{sat}\longrightarrow \infty$.}
\label{fig:satper}
\end{figure}

In our next experiment, we have analyzed the performance of LSM-DER with the variation of $x_{sat}$ (defined in Section \ref{lsm_der}). Fig.\ref{fig:satper}(a) and (b) demonstrates the dependence of MAE on $x_{sat}$ for tasks I and II respectively. As expected, small values of $x_{sat}$ lead to higher error since the branch outputs saturate and cannot encode changes in input. The ideal situation, like \cite{Mel2001}, is defined when there is no $x_{sat}$ i.e. the dendritic branches produce a square law output without an upper bound and is denoted by the dashed lines in the figure.

\begin{figure}[!t]
	\includegraphics[width=0.45\textwidth]{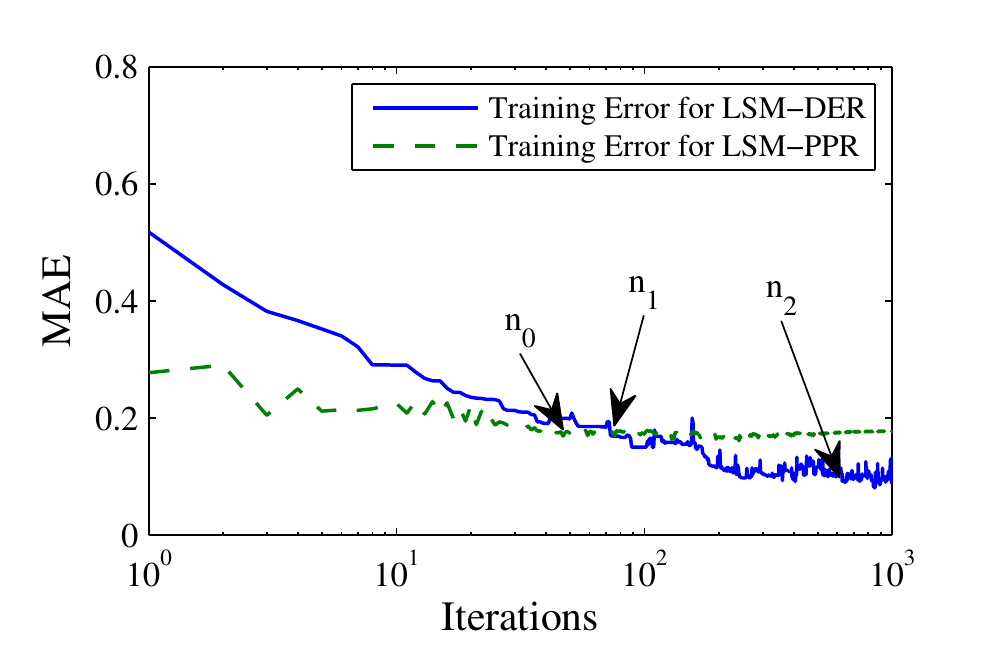}
\caption{\small MAE vs iterations curve of LSM-DER and LSM-PPR for $200$ input patterns averaged over $10$ trials for Task I.}
\label{fig:maeiter}
\end{figure}

\subsection{Results: Comparison between LSM-DER and LSM-PPR}
Next, we compare the performance of LSM-DER with LSM-PPR. In Fig.\ref{fig:maeiter}, the profiles of MAE during the training procedure associated with Task I is shown for LSM-DER and LSM-PPR; the results and conclusions for Task II are qualitatively similar. It is evident from Fig.\ref{fig:maeiter} that during Task I, the training error for LSM-PPR reduces swiftly and saturates before LSM-DER converges to its minimum error. On the other hand, LSM-DER takes more number of iterations to reach the minimum error but the minimum training error obtained by it is less than that of LSM-PPR. Thus, from the two curves of Fig.\ref{fig:maeiter} we can mark three significant points:

\begin{enumerate}
  \item $n_0$: Number of iterations at which the error for LSM-PPR saturates.
  \item $n_1$: Number of iterations at which the error curve of LSM-DER and LSM-PPR intersects for the first time.
  \item $n_2$: Number of iterations required by LSM-DER to achieve minimum error.
\end{enumerate}

\begin{figure}
\begin{center}
\includegraphics[width=0.45\textwidth]{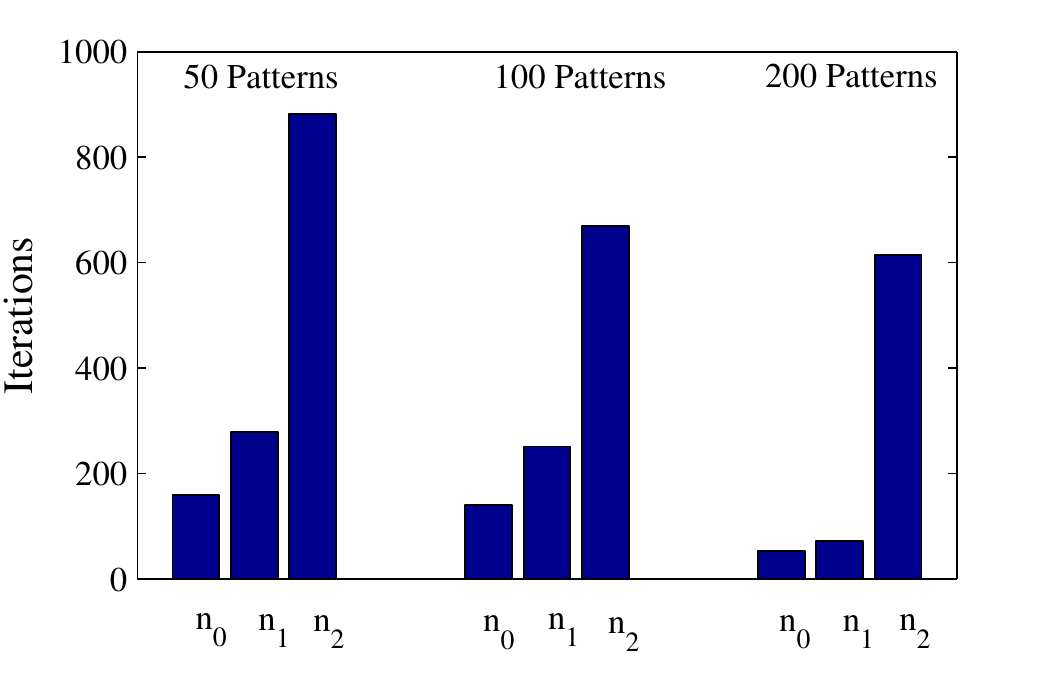}
\caption{\small Convergence Analysis of LSM-DER and LSM-PPR for Task I: The average value of $n_0$, $n_1$ and $n_2$ obtained for $50$, $100$ and $200$ input patterns averaged over 10 trials.   \label{fig:conv_ana}}
\end{center}
\end{figure}

We have also compared the convergence in training for LSM-DER and LSM-PPR when the number of input patterns are $50$, $100$ and $200$. The average values $n_0$, $n_1$ and $n_2$ obtained for the three cases are depicted in Fig.\ref{fig:conv_ana} for Task I. From this figure, we can see that $n_0<n_2$ for all cases implying faster convergence for LSM-PPR. However, we can note that $n_1\approx n_2$ implying LSM-DER can always achieve same error as LSM-PPR  in roughly similar number of iterations; it takes more time for LSM-DER to find \emph{better} solutions.

Till now, for our simulations the number of liquid neurons $L$ was kept as $140$. Now, we will look into the effect of increasing liquid size on the performance and convergence of LSM-DER and LSM-PPR for Task I. For $L=560$ LSM-DER provides a testing error of $0.079$ and $n_2=1472$. On the other hand for $L=560$, LSM-PPR provides a testing error of $0.129$ and the average number of iterations it takes to saturate while training is ($n_0$) is $628$. For $L=1120$ LSM-DER and LSM-PPR provides testing error of $0.076$ and $0.12$ respectively. In this case, $n_2$ for LSM-DER is $2402$ and $n_0$ for LSM-PPR is $914$. Thus we see that for higher dimensional inputs, the NRW rule is still able to find suitable connections but requires a larger number of iterations.

\begin{figure}[!t]
\centering
   \subfloat[]{\includegraphics[width=0.25\textwidth]{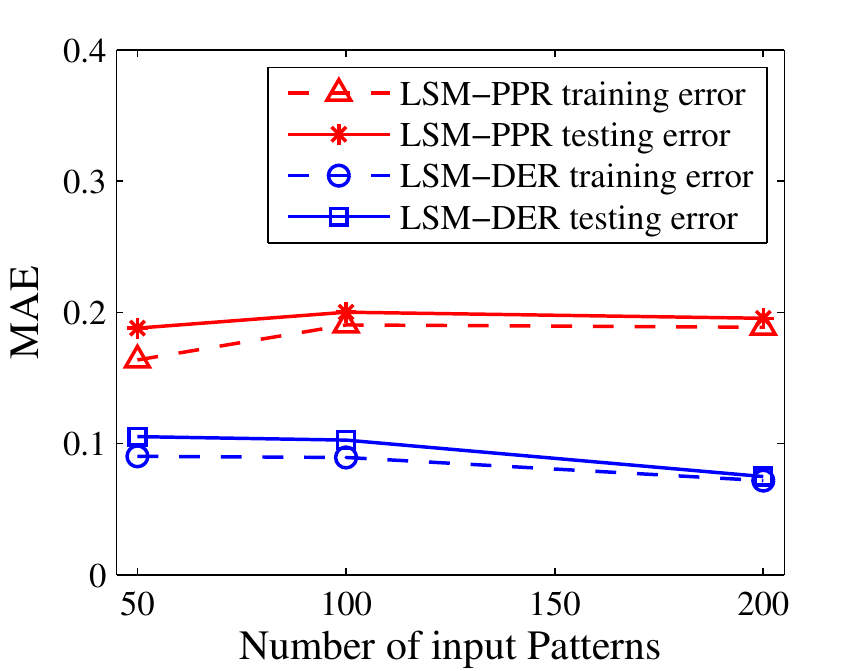}}
   \subfloat[]{\includegraphics[width=0.25\textwidth]{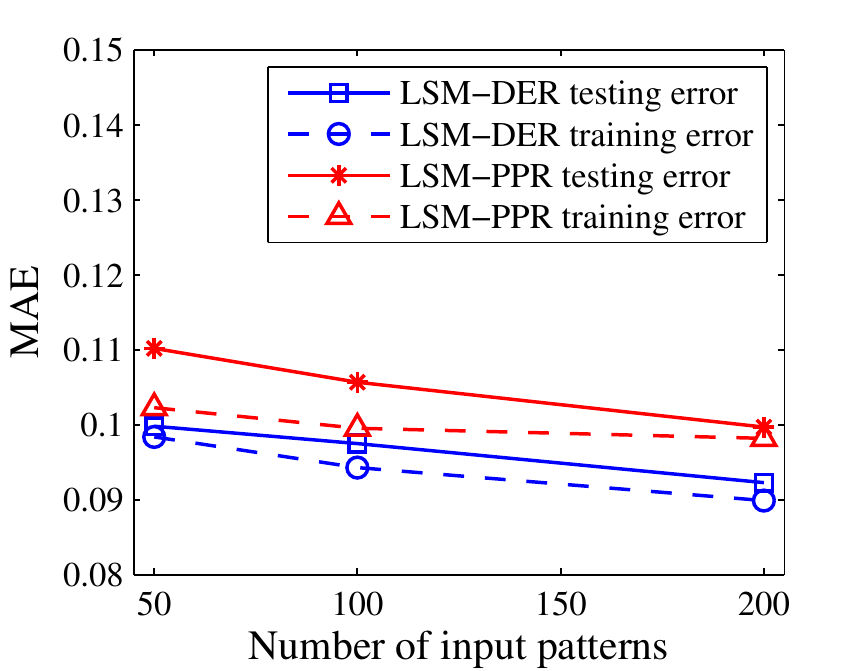}}
\caption{\small Generalization Plot: A performance comparison of the training and testing error of LSM-DER and LSM-PPR for $50$, $100$ and $200$ input patterns averaged over $10$ trials on (a)Task I and (b) Task II.}
\label{fig:gen_plot}
\end{figure}

The earlier plots suggest LSM-DER can attain better error in training--however, error during testing is more important to show the ability of the system to generalize. These generalization plots depicted in Fig.\ref{fig:gen_plot}(a) and (b) for Tasks I and II respectively show that for each case, LSM-DER outperforms LSM-PPR for different number of training patterns. 
	 
\begin{figure}[!t]
\centering
   \subfloat[]{\includegraphics[width=0.45\textwidth,height=4.5cm]{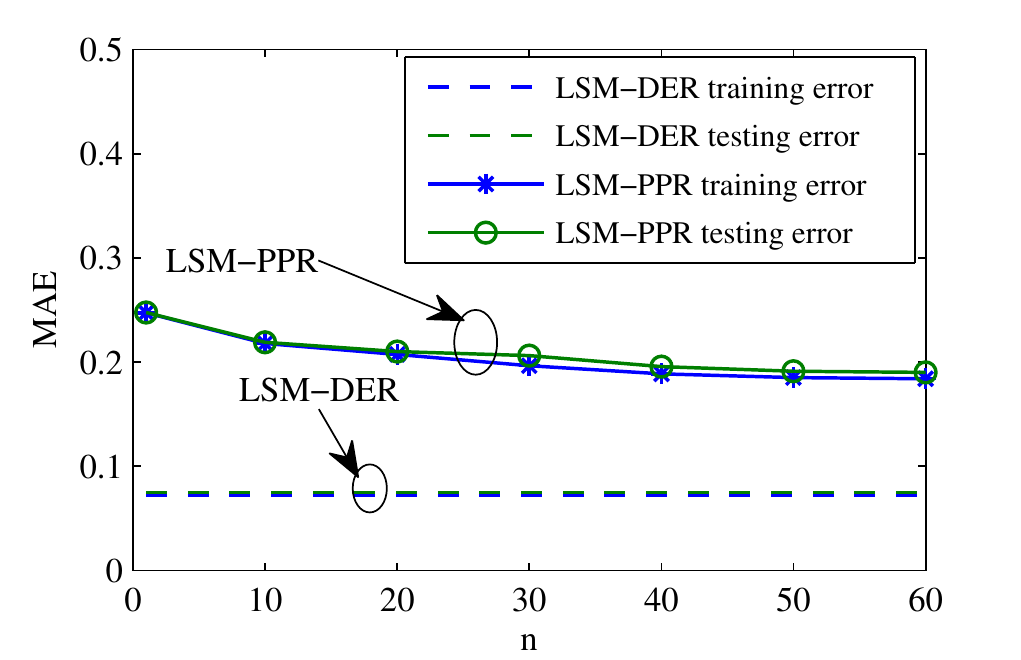}}\\
   \subfloat[]{\includegraphics[width=0.45\textwidth,height=4.5cm]{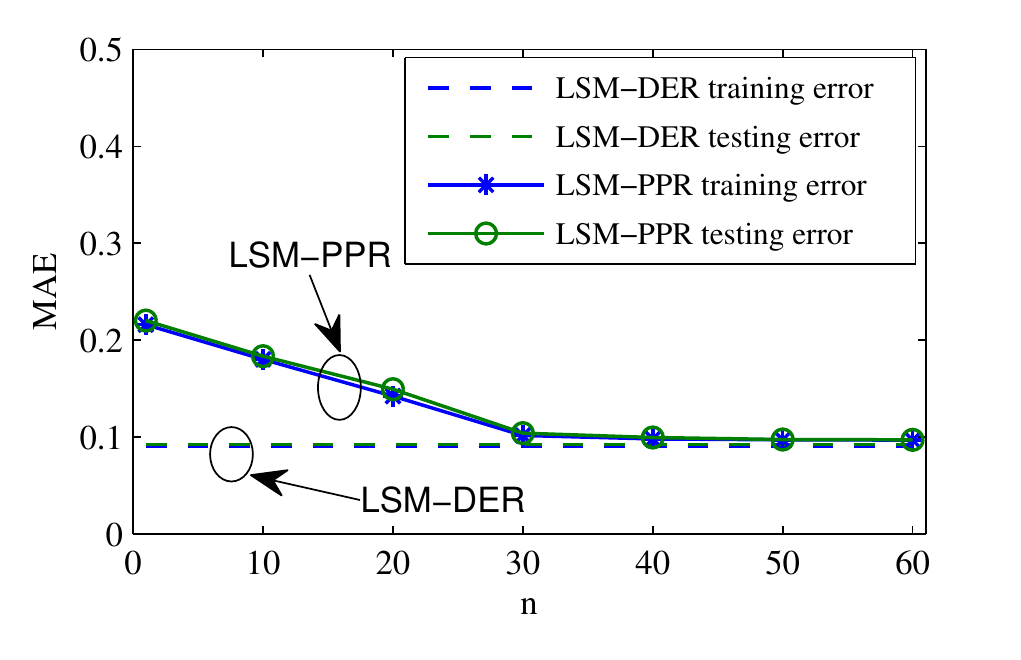}}
\caption{Performance Comparison of LSM-DER and LSM-PPR with varying $n$ for (a)Task I and (b) Task II: The classification error gradually decreases with increasing $n$ and finally it saturates to a value higher than the error attained by LSM-DER.}
\label{fig:pvar}
\end{figure}

LSM-PPR also requires the setting of another parameter: $n$ denoting the number of readout neurons. Till now, LSM-DER has been compared with LSM-PPR keeping $n=40$. Next, we vary the number of readout neurons of LSM-PPR and compare the results with LSM-DER. The outcome of these experiments are shown in Fig.\ref{fig:pvar}(a) and (b). In both the figures, the training and testing error of LSM-DER is plotted as a constant line and is compared with the results of LSM-PPR for different values of $n$. During this experiment, we chose the values of $n$ as 1, 10, 20, 30, 40, 50 and 60 thereby covering the most basic $n=1$ case to advanced cases employing more number of readout neurons. Note that the synaptic resource consumed by LSM-DER is same as the case for $n=1$ or a single perceptron. As evident from the figures, the error attained by LSM-PPR decreases with the increase in value of $n$ but finally becomes saturated at a value higher than the error attained by LSM-DER. The graph also shows that the saturation starts approximately when $n=40$ and this also explains why we have chosen this value of $n$ while comparing our LSM-DER algorithm to LSM-PPR.From these plots, we can conclude that LSM-DER attains $3.3$ and $2.4$ times less error than LSM-PPR with same number of high resolution weights in tasks I and II respectively. Also, LSM-PPR requires $40-60$ times more synapses to achieve similar performance as LSM-DER in Task II.

\section{Discussion}
\label{dis}
\begin{figure}
\begin{center}
\includegraphics[width=0.35\textwidth]{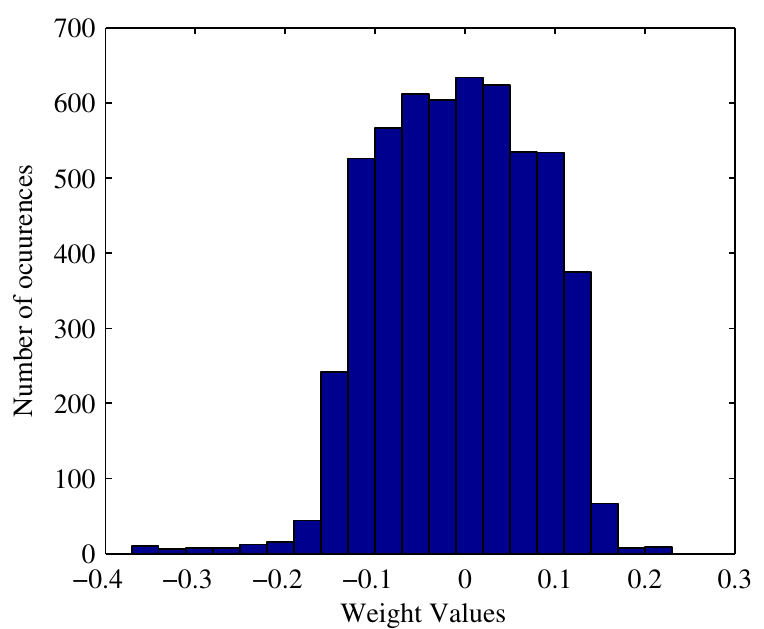}
\caption{\small Histogram showing the distribution of weights of the synapses in the readout for LSM-PPR. It can be noted that most of the synapses have a non-zero weight implying that most of the synapses are active and add to resource consumption. \label{fig:histo}}
\end{center}
\end{figure}
It is surprising that even after using $140\times 40$ synapses, LSM-PPR cannot attain the performance of LSM-DER that utilizes only $140$ synapses. To verify if this is indeed an improvement, we need to check if LSM-PPR is really using up all the allocated synapses. Thus we provide a histogram in Fig.\ref{fig:histo} which shows the final weight distribution of the synapses for Task I when trained on $200$ patterns. From the figure it is clear that most of the weights are non-zero thereby confirming that a large portion of the $140\times 40$ synapses are indeed active and being used. This is surprising since the function counting method (used to predict the capacity of the NL-cell) would predict a large capacity for the parallel perceptron case as well. Essentially, each of the perceptrons behave akin to a dendrite--in fact, they have more number of analog weights and hence could have better capacity. The differences between the DER and PPR readout are in the threshold nonlinearity and the training algorithm. Also, the sampled inputs $s_a'(t)$ for this stage are derived through a different convolution kernel. To tease apart the contribution of each difference to the poor performance, we consider two separate cases for training LSM-PPR on Task I. In case I, we consider PPR readout with same number of perceptrons as that of dendrites used by DER i.e. we take $n=2m=14$ but instead of the threshold nonlinearity, we use a saturating square non-linearity similar to LSM-DER. In this case, LSM-PPR is able to reduce the average MAE from $0.216$ to $0.133$ indicating the advantage of preserving some analog information at the output of each perceptron. In case II, we explore the impact of different states $s_a'(t)$. We keep $n=1$ for LSM-PPR, as for this case both the algorithms use same number of synaptic resources ($s=140$) for connecting the liquid and the readout, and use our convolution kernel instead of the one in the LSM toolbox. In this case,  the average MAE reduces to $0.136$ showing the importance of choosing the kernel carefully. Next, combining both these modification, LSM-PPR with $n=2m=14$ perceptrons and the new kernel function could be trained by \emph{p}-delta to achieve an average MAE of $0.071$ that is comparable to LSM-DER.

\begin{figure}[!t]
\centering
   \subfloat[]{\includegraphics[width=0.45\textwidth]{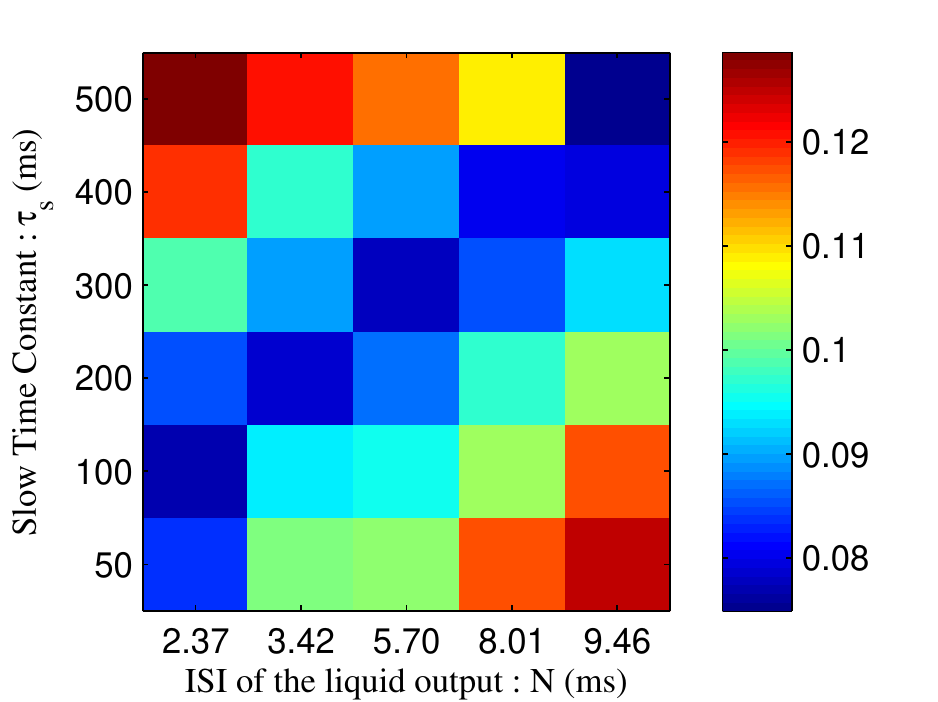}}\\
   \subfloat[]{\includegraphics[width=0.45\textwidth]{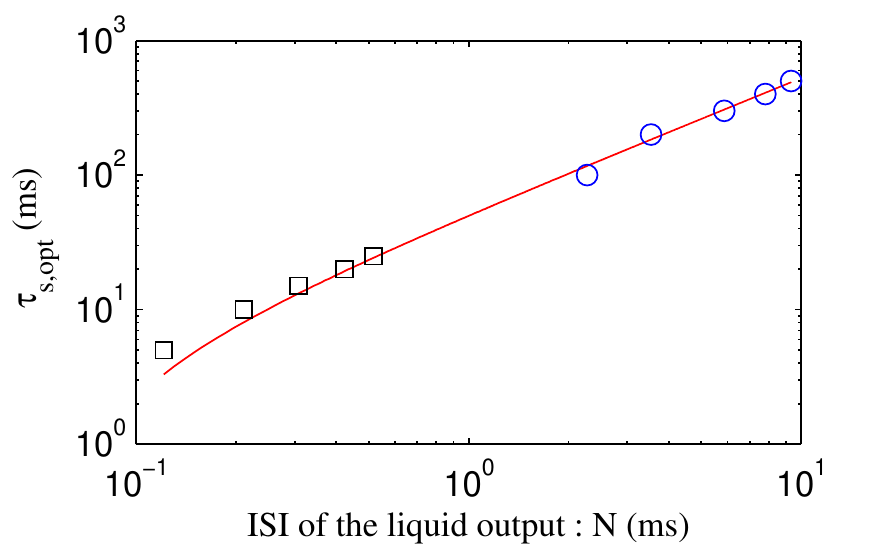}}
\caption{\small (a) MAE for Task I is reported by varying ${\tau_s}$ and inter spike interval of the liquid output  $N$. (b) A curve fit to the optimal ${\tau_s}$ for each $N$ is obtained by varying $N$ for both tasks over two orders of magnitude.}
\label{fig:tau_var_L}
\end{figure}

Given the importance of the convolution kernel for generating $s_a'(t)$ as described earlier, we now provide some insight to its choice. The state generation method discussed in Section \ref {sec:spike_gen} requires two parameters : $\tau_s$ and $\tau_f$. $\tau_f$ is the fast time constant which takes a small positive value in hardware implementations and is typically not tuned. $\tau_s$ is the slow time constant responsible for integration across temporally correlated spikes and we will analyze its effect on the performance of LSM-DER. If there are $L$ liquid neurons and the mean firing rate of the each liquid neuron is $\mu_f$, then the mean ISI across the entire liquid output is given by $N=1/ (L \times \mu_f)$. Intuitively, we expect $\tau_{s,opt}$, the optimal value of $\tau_s$, to be correlated with this quantity since it has to be long enough to integrate information of temporally correlated spikes across all the liquid neurons. In Fig.\ref{fig:tau_var_L}(a), the MAE attained by LSM-DER for Task I is shown for different values of $\tau_s$ when $N$ is varied by changing the value of $L$. A similar result is obtained by changing the value of $\mu_f$ keeping $L$ as constant and is not shown here to avoid repetition. From this figure, we see a strong correlation between $N$ and the best value of $\tau_s$, $\tau_{s,opt}$. A similar set of simulations are also done for Task II. The best $\tau_s$ is then chosen for each value of $N$ and plotted in Fig.\ref{fig:tau_var_L}(b). It can be seen that for both tasks and with $N$ spanning two orders of magnitude, $\tau_{s,opt}$ can be well described by the fit $\tau_{s,opt}=52.83N-3.1$.

\section{VLSI Implementation: Effect of Statistical Variations}
This section contains the description of a  VLSI architecture to implement DER and PPR, Monte Carlo simulation results of the key sub-circuits to show their statistical variability and incorporation of these statistical variations in Matlab analytical models to analyze the stability of the algorithms. These circuits are designed in AMS $0.35$ $\mu$m CMOS technology.

\begin{figure*}[ht]
\begin{center}
\includegraphics[width=0.9\textwidth]{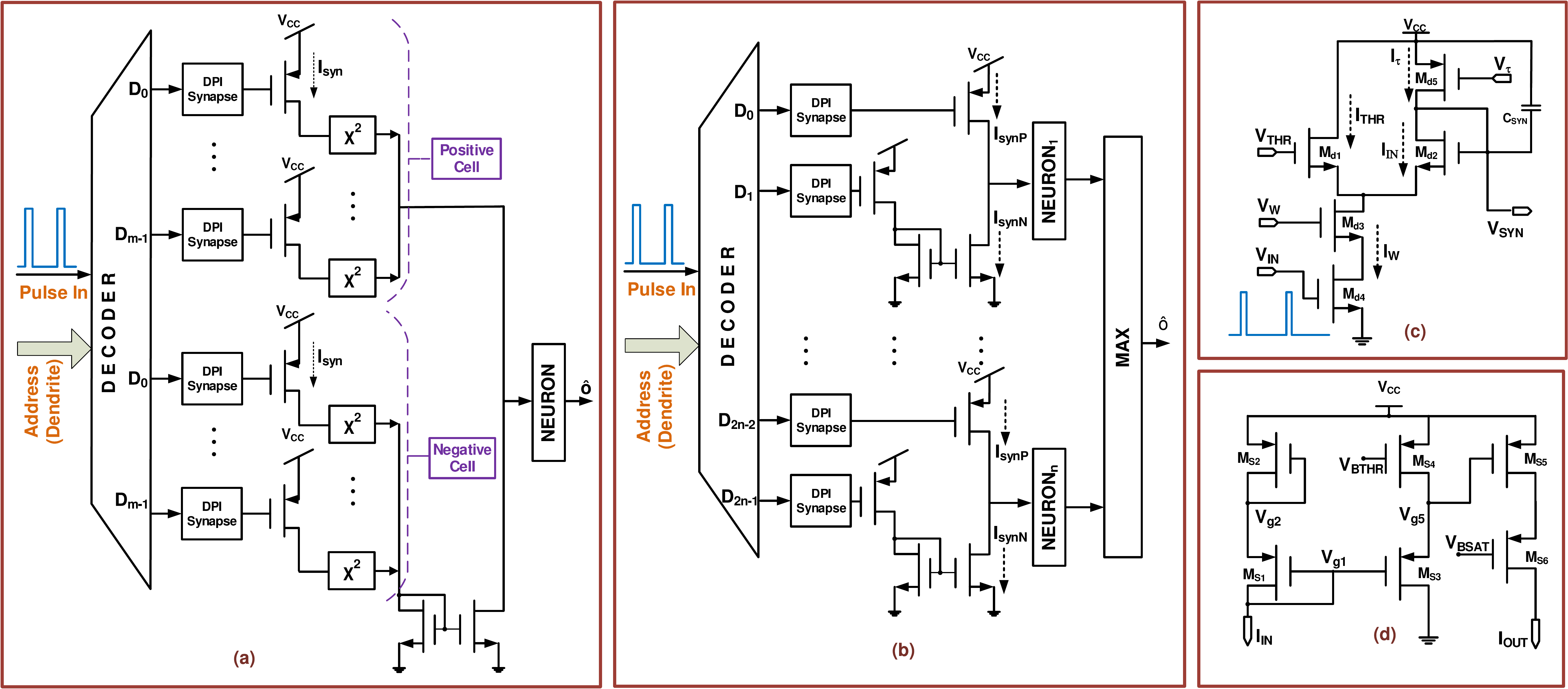}
\caption{Architecture for VLSI implementation of (a) DER and (b) PPR readout schemes for LSM. Schematic diagram of DPI Synapse and squaring block used in these architectures are shown in detail in (c) and (d) respectively.}
 \label{fig:ArchitectureCircuitsNeuron_Diagram}
\end{center}
\end{figure*}

\subsection{VLSI Architecture of the Spiking Neural Network}
The VLSI architectures for implementing DER and PPR readouts for LSM are shown in Fig. \ref{fig:ArchitectureCircuitsNeuron_Diagram}(a) and (b) respectively. For both cases, AER is used to provide the synaptic connections. For DER, there is one shared synapse for every dendritic branch while for PPR, there is one shared synapse per perceptron. The input spikes (output of the liquid) are applied to the circuit through an address decoder while Differential Pair Integrator (DPI) circuits are used to implement synaptic function. For one neuron of the DER case, there are $m$ dendritic branches connected to a NEURON block through $m$ Square Law Nonlinear circuits. The output of the spiking neuron can be converted to the analog output $y$ by considering the spike rate averaged over a pre-defined time period. For the case of classification, a winner-take-all circuit can be used instead of two neurons. The VLSI architecture for PPR with $n$ perceptrons is the combination of $n$ DPI circuits with current source outputs (for positive weights) and $n$ with current sink outputs (for negative weights). For each perceptron, if the current from the positive weight DPI is higher, the output voltage gets pulled high and vice versa. These voltages are the inputs to the MAX block that generates the output decision by voting. Compared to the NEURON or MAX blocks, the synapses and square law blocks are more numerous. Hence, their individual sizes are also kept small leading to them being more susceptible to variations. Next, we shall briefly describe circuit implementations of these blocks and show simulations of the effect of statistical variations.

\subsubsection{Differential Pair Integrator Synapse}
The circuit schematic shown in Fig. \ref{fig:ArchitectureCircuitsNeuron_Diagram}(c) is a DPI circuit that converts presynaptic voltage pulses into postsynaptic currents ($I_{syn}(t)$). In this circuit, transistors operate in subthreshold regime. As mentioned in \cite{Bartolozzi_Indiveri_SynapticDynamics}, controlling the bias voltages $V_w$ and $V_{tau}$ we can set $I_w$$\gg$$I_{\tau}$, which simplifies this nonlinear circuit to a canonical first-order low pass filter. The bias voltages $V_{thr}$, $V_w$ can effectively control the weight of the synapse (maximum output synaptic current, denoted by $I_0$) and $V_{tau}$ controls the fall time constant ($\tau_s$) of the output current. For an input spike arriving at $t_i^-$ and ending at $t_i^+$ to the DPI synapse, the rise time of $I_{syn}$ is very small. The discharge profile can be modeled by the Equation \ref{eqn:SynapseBlockEquation}.

\begin{equation}
    \label{eqn:SynapseBlockEquation}
        \begin{array}{lcl}
         I_{syn}(t)=I_0e^{(-\frac{t-t_i^+}{\tau_s})}
        \end{array}
\end{equation}

where $\tau_s$=$\frac{C_{SYN}U_T}{\kappa I_\tau}$, $\kappa$ is the subthreshold slope factor, and $U_T$ is the thermal voltage.

\begin{table}[!ht]
  \caption{Monte Carlo Simulation Results of DPI Synapse}
  \centering
  \begin{tabular}{|c|c|c|c|c|c|c|}
  \hline
    $\mu[I_0]$ & $\sigma[I_0]$  &  $\frac{\sigma}{\mu}[I_0]$    & $\mu[\tau_s]$ & $\sigma[\tau_s]$  &  $\frac{\sigma}{\mu}[\tau_s]$\\ \hline
   4.21 nA    &  570 pA      &   13 \%    &  912.5 $\mu s$   & 91.9 $\mu s$   & 10.1 \% \\
    1.06 nA    &  127 pA      &   12 \%    &  903.1 $\mu s$   & 90.7 $\mu s$   & 10.0 \% \\
    7.74 nA    &  875 pA      &   11 \%    &  4116 $\mu s$    & 413.6 $\mu s$  & 10.1 \% \\
    1.06 nA    &  127 pA      &   12 \%    & 18335.7 $\mu s$  & 1832.7 $\mu s$ & 10.0 \% \\
  \hline
  \end{tabular}
  \label{table:MCsimulationResultsDPI}
\end{table}

\subsubsection{Saturating Square Law Nonlinear circuit}
We have designed the current mode squaring circuit given in Fig. \ref{fig:ArchitectureCircuitsNeuron_Diagram}(d) as described in \cite{shaista_ijcnn1}. Transistors M$_{S2}$, M$_{S1}$, M$_{S3}$ and M$_{S5}$ form a translinear loop. Hence, the current through M$_{S5}$ is expressed as given in Equation \ref{eqn:SquareBlockEquation}. The transistor M$_{S5}$ is biased to pass a maximum current of $I_{sat}$ (set by $V_{BSAT}$).

\begin{equation}
    \label{eqn:SquareBlockEquation}
    I_{out}=\frac{I_{in}^2}{I_{thr}}
\end{equation}

$I_{thr}$ is the dc current through M$_{S4}$ set by its Gate voltage ($V_{BTHR}$). However, due to process parameter mismatch between the transistors M$_{S2}$, M$_{S1}$, M$_{S3}$ and M$_{S5}$, the output current deviates from the exact relationship given in Equation \ref{eqn:SquareBlockEquation}. Since variation of threshold voltage ($\bigtriangleup V_{th}$) dominates other sources of variation in the subthreshold regime, the translinear loop equation for Fig. \ref{fig:ArchitectureCircuitsNeuron_Diagram}(d) can be re-written as:

\begin{align}
    \label{eqn:SquareBlockMismatchEquation}
    I_{out}^\prime&=e^{\frac{(\bigtriangleup V_{th1}+\bigtriangleup V_{th2}-\bigtriangleup V_{th5}-\bigtriangleup V_{th3})}{U_T}} \times \frac{I_{in}^2}{I_{thr}}\notag\\
    &=c_{ni}\times I_{out}
\end{align}

where $I_{out}^\prime$ is the actual current, $I_{out}$ is the expected current without mismatch and $c_{ni}=e^{\frac{(\bigtriangleup V_{th1}+\bigtriangleup V_{th2}-\bigtriangleup V_{th5}-\bigtriangleup V_{th3})}{U_T}}$ models the nonideality term due to mismatch.

\subsection{Monte Carlo Simulation Results}
We have performed Monte-Carlo simulation of the DPI synapse and Square Law Nonlinear circuits considering transistor mismatch. The objective of the Monte-Carlo simulation is to capture the variation of $I_0$, $\tau_s$, $c_{ni}$ from one dendritic branch (or perceptron) to the other. Some of the representative results of the statistical simulation are listed in Table \ref{table:MCsimulationResultsDPI} and \ref{table:MCsimulationResultsSquareBlock}. For different settings of variable bias parameters ($V_{thr}$, $V_{tau}$, $V_w$ and $V_{BTHR}$), we obtained worst case variation of $I_0$ and $I_{out}^\prime$ as $13\%$ and $18\%$ respectively. The parameter $\frac{\sigma}{\mu}[I_{OUT}^\prime]$ given in Table \ref{table:MCsimulationResultsSquareBlock} can also be written as $\frac{\sigma}{\mu}[c_{ni}]$.

\begin{table}[!ht]
\caption{Monte Carlo Simulation Results of Square Law circuit}
\centering
\begin{tabular}{|p{0.9cm}|p{0.9cm}|p{1.1cm}|p{1.1cm}|p{2.5cm}|}
  \hline
  $I_{thr}$ & $I_{IN}$   & $\mu[I_{OUT}^\prime]$ & $\sigma[I_{OUT}^\prime]$  &  $\frac{\sigma}{\mu}[I_{OUT}^\prime]$ = $\frac{\sigma}{\mu}[c_{ni}]$  \\ \hline
  30 nA   & 20 nA       & 12.15 nA     & 2.20 nA     &  18.0 \% \\
  30 nA   & 60 nA       & 99.70 nA     & 13.6 pA     &  14.0 \% \\
  30 nA   & 100 nA      & 243.1 nA       & 26.4 pA     &  11.1 \% \\
  \hline
\end{tabular}
\label{table:MCsimulationResultsSquareBlock}
\end{table}
\begin{figure}[t]
\centering
   \subfloat[]{\includegraphics[width=0.25\textwidth]{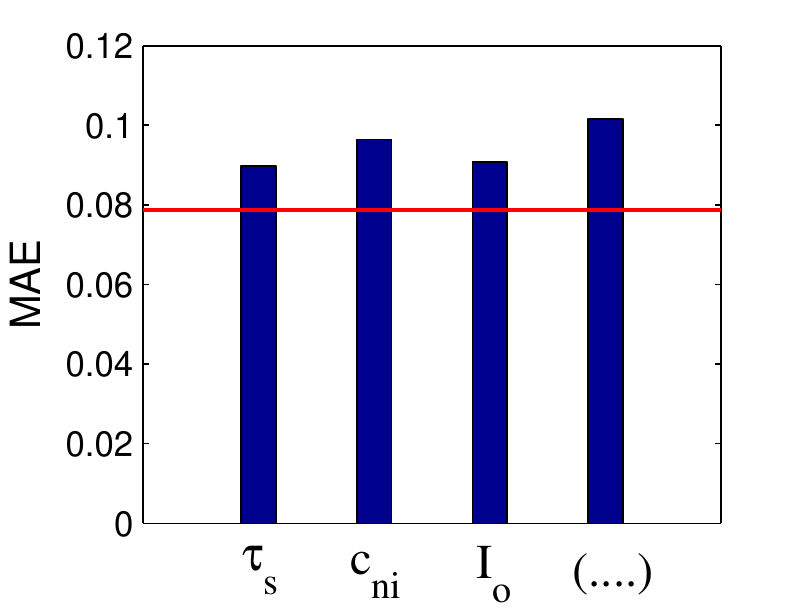}}
   \subfloat[]{\includegraphics[width=0.25\textwidth]{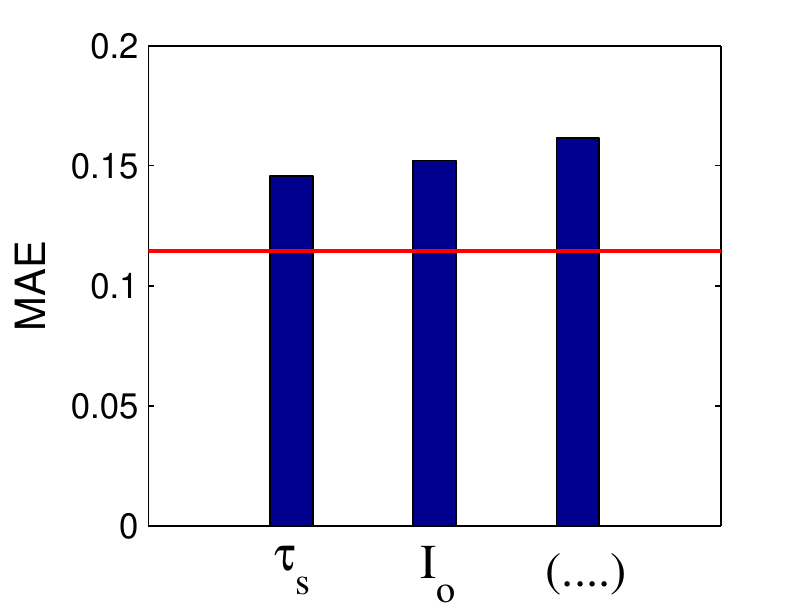}}
\caption{Stability of LSM-DER and LSM-PPR with respect to different hardware non-idealities are plotted for Task I in (a) and (b) respectively. The constant red line indicates the testing error obtained by the algorithms without any non-ideality. See text for details.}
\label{non_ideal}
\end{figure}
The impact of global process variation on circuit performance has been neglected here, because performance drift due to global variations can be eliminated by tuning the bias parameters. The device sizing for the circuit blocks are given in Table \ref{table:WbyL_sizingTransistors} for reference.

\begin{table}[!ht]
\caption{Design parameters of the Circuit Blocks in Fig. \ref{fig:ArchitectureCircuitsNeuron_Diagram}}
\centering
\begin{tabular}{|c|c|}
  \hline
  Synapse Block in Fig. \ref{fig:ArchitectureCircuitsNeuron_Diagram} (c) &  Square Block in Fig. \ref{fig:ArchitectureCircuitsNeuron_Diagram} (d) \\ \hline
  $\frac{W}{L}$ ($M_{d1}$, $M_{d2}$, $M_{d3}$) $\equiv$ $\frac{8 \mu}{1 \mu}$,   & $\frac{W}{L}$ ($M_{s1}$, $M_{s2}$) $\equiv$ $\frac{8 \mu}{2 \mu}$,      \\
  $\frac{W}{L}$ ($M_{d4}$) $\equiv$ $\frac{0.40 \mu}{0.35 \mu}$, $\frac{W}{L}$ ($M_{d5}$) $\equiv$ $\frac{10 \mu}{1 \mu}$   &   $\frac{W}{L}$ ($M_{s3}$, $M_{s5}$) $\equiv$ $\frac{8 \mu}{2 \mu}$,    \\
  $C_{syn}=1.1 pF$   & $\frac{W}{L}$ ($M_{s4}$, $M_{s6}$) $\equiv$ $\frac{16 \mu}{2 \mu}$     \\
  \hline
\end{tabular}
\label{table:WbyL_sizingTransistors}
\end{table}

\subsection{Area comparison between LSM-DER and LSM-PPR}
Since the liquid area would be same for both LSM-DER and LSM-PPR, we will only concentrate on the area of the readout. Let us denote the areas of a neuron, dendritic non-linear square block and synapseby $A_{neu}$, $A_{den}$ and $A_{syn}$ respectively. Then, DER would have an area of $A_{DER}=A_{neu}+2m(A_{den}+A_{syn})$ whereas PPR would have an area of $A_{PPR}=n (A_{neu}+ 2A_{syn})$, where $n$ is the number of perceptrons in DER. In our VLSI implementation, $A_{neu}=68 \mu m \times 38 \mu m$, $A_{den}=75 \mu m \times 28 \mu m $ and $A_{syn}=75 \mu m \times 42 \mu m$. Considering $m=7$ and $n=40$, $A_{DER}=76084 (\mu m)^2$ and $A_{PPR}=355360(\mu m)^2$ i.e. $A_{PPR}\approx 4.67 A_{DER}$.

\subsection{Stability Analysis in Software Simulations}\label{stab_ana}
To analyze the stability of the algorithms, the statistical variations described above are incorporated during the testing phase of the simulation. The non-idealities are included only in the testing phase (and not during the training phase) because in the actual implementation, the training will be done in software and the trained connections will be downloaded directly to the chip.  Fig.\ref{non_ideal} shows the performance of both LSM-DER and LSM-PPR when the non-idealities are included for Task I. Since all the variations are across branch, for fair comparison the PPR architecture used in this analysis has the same number of perceptrons as that of the dendrites used by LSM-DER.  Moreover, we have used the same convolution kernel for state generation in both cases. In Fig. \ref{non_ideal}, the bars corresponding to $\tau_s$, $c_{ni}$, and $I_0$ denote the performance degradation when statistical variations of $\tau_s$, $c_{ni}$ and $I_0$ are included individually. Finally, to imitate the true scenario we consider the simultaneous implementations of all the non-idealities, which is marked by $(...)$. In the software simulations, the $\frac{\sigma}{\mu}$ of $\tau_s$ and $I_0$ has been taken to be the worst case scenario as displayed in Table \ref{table:MCsimulationResultsDPI} i.e. $10.1\%$ and $13\%$ respectively. Similarly, the $\frac{\sigma}{\mu}$  of $c_{ni}$ in the software simulations has been taken to be the worst case scenario as portrayed in Table \ref{table:MCsimulationResultsSquareBlock} i.e. $18\%$. From Fig. \ref{non_ideal} it is evident that when all the variations are included ,then the MAE of LSM-DER and LSM-PPR increases by $0.0233$ and $0.0470$ respectively. This concludes that the modifying connections of binary synapses in LSM-DER results in more robust VLSI implementations compared to the adaptation of high resolution weights in LSM-PPR.

\section{Conclusion}
In this article we have proposed a novel architecture (LSM-DER) and an efficient learning rule (NRW) for the readout stage of Liquid State Machine. Inspired by the nonlinear properties of dendrites in biological neurons, the readout neurons of LSM-DER employs multiple dendrites with lumped nonlinearities. The results depict that the advantages of LSM-DER along with NRW over the state-of-the-art LSM-PPR \cite{Maass2002} are: 
\begin{itemize}
	\item The LSM-DER algorithm attains less error than LSM-PPR for both classification and approximation problems as shown in detail in Section III. 
	\item If there are $L$ liquid neurons, then LSM-PPR required $L\times n$ synapses for connecting the liquid to the readout whereas LSM-DER can achieve comparable performance with far fewer synapses. Moreover, when same number of synapses are allocated to both LSM-DER and LSM-PPR, LSM-DER achieves $3.3X$ less error in classification and $2.4X$ less error in approximation.
	\item LSM-PPR requires analog synaptic weights whereas LSM-DER can achieve better performance even with binary synapses, thus being very advantageous for hardware implementations. Since the synapses are binary valued, the NRW learning rule chooses the best possible connection matrix between inputs and the dendritic branches. 
\end{itemize}
Also, we have shown that the proposed architecture is more robust against statistical variation of parameters than LSM-PPR, a feature essential for VLSI implementations.

\bibliographystyle{IEEEbib}

\end{document}